\documentclass[a4paper,11pt]{article}
\pdfoutput=1
\usepackage{jheppub}
\usepackage[ascii]{inputenc}
\usepackage{bbm,braket,microtype,mathtools}
\usepackage{amsmath}
\usepackage[T1]{fontenc}
\usepackage{lmodern}
\usepackage[USenglish]{babel}
\DeclareMathOperator{\tr}{tr}

\allowdisplaybreaks[4]

\definecolor{pink1}{rgb}{0.858, 0.188, 0.478}
\definecolor{red1}{RGB}{213,4,4}
\definecolor{blue1}{RGB}{0,179,255}

\def\text#1{{\rm #1}}

\def\avg#1{\left\langle#1\right\rangle}
\def\bra#1{\left\langle#1\right|}
\def\ket#1{\left|#1\right\rangle}

\def\abs#1{\left|#1\right|}
\def\kc#1{\left(#1\right)}
\def\kd#1{\left[#1\right]}
\def\ke#1{\left\{#1\right\}}

\def\trace#1{{\rm Tr}\left[#1\right]}

\def\be{\begin{equation}}       \def\ee{\end{equation}}
\def\bea{\begin{eqnarray}}      \def\eea{\end{eqnarray}}
\def\ba{\begin{array} }
\def\ea{\end{array} }

\def\nn{\nonumber}
\def\pa{\partial}
\def\=>{\Rightarrow}
\def\>{\rightarrow}

\def\addfig#1#2#3{\begin{figure}
\includegraphics[width=#1]{#2}
\caption{#3
}
\end{figure}}

\begin{document}

\title{Space-time random tensor networks and holographic duality}

\author{Xiao-Liang Qi,}
\author{and Zhao Yang}

\affiliation{Stanford Institute for Theoretical Physics, \\
Physics Department, Stanford University, CA 94304-4060, USA}

\abstract{In this paper we propose a space-time random tensor network approach for understanding holographic duality. Using tensor networks with random link projections, we define boundary theories with interesting holographic properties, such as the Renyi entropies satisfying the covariant Hubeny-Rangamani-Takayanagi formula, and operator correspondence with local reconstruction properties. We also investigate the unitarity of boundary theory in spacetime geometries with Lorenzian signature. Compared with the spatial random tensor networks, the space-time generalization does not require a particular time slicing, and provides a more covariant family of microscopic models that may help us to understand holographic duality. 
}

\maketitle

\section{Introduction}


Holographic duality proposes that a $d$ dimensional quantum field
theory can be equivalently described by a $d+1$ dimensional quantum
gravity\cite{maldacena1998,witten1998,gubser1998}. The role of quantum
entanglement in the holographic duality manifests itself in the
Ryu-Takayanagi (RT) formula \cite{ryu2006holographic} and its
generalizations\cite{hubeny2007covariant,faulkner2013quantum,lewkowycz2013generalized,dong2016deriving,dong2016gravity}. The
RT formula and other properties of the holographic duality have
motivated the tensor network approaches as an effort to develop a
microscopic framework for holographic duality\cite{swingle2012,qi2013,
  pastawski2015, yang2015, hayden2016, qi2017holographic,
  donnelly2016living}. Most tensor networks define states on a time
slice. In appropriate limits they reproduce the RT formula, where the
bulk geometry is replaced by the graph geometry in the tensor network
approach.  Thus these tensor networks resemble the static geometry
({\it i.e.} a geometry with a time-like Killing vector) where the RT
formula applies. However, for more general dynamical geometries, the
minimal surface and RT formula is not well-defined, and the
entanglement entropy of a boundary region is dual to the area of an
extremal co-dimension-$2$ surface in the bulk space-time, known as the
Hubeny-Rangamani-Takayanagi (HRT)
formula\cite{hubeny2007covariant}. 
In the (spatial) tensor network picture, it is often imagined that the
dynamical space-time is described by a spatial tensor network with a
time-dependent geometry. However, there is an intrinsic problem with
this picture.  In contrast to the case of static space-time, in
general, HRT surfaces of different boundary regions at the same
boundary time cannot be embedded into a single Cauchy
surface. Therefore it is impossible to find a ``proper" Cauchy surface
and describe the boundary state as a tensor network satisfying RT
formula on this surface.
Besides, the spatial tensor network description requires the choice of a time direction, which can not manifest the general covariance.  Motivated by these problems, we develop a more covariant space-time tensor network description to holographic duality.

In this paper, we propose a new approach to the holographic duality
based on space-time tensor networks. The tensor network we consider is
defined in the bulk space-time, with random projections applied to
each link (more details will be described later). Instead of
describing a many-body wavefunction, the space-time tensor network
defines the boundary parition function with arbitrary insertions. In
other words, it defines the generating function of all (time-ordered)
multi-point functions of the boundary theory. A time-ordered
multi-point function is simply obtained by insertion of operators in
the boundary links of the tensor network. We show that the Renyi
entropy of a boundary region, after averaging over random projections
in the bulk, can be mapped to the partition function of a discrete
gauge theory. In the large bond dimension limit the gauge theory is in
the classical limit, and the Renyi entropy is determined by the
classical energy of the minimal action gauge field configuration. With
the boundary condition determined by the boundary region, we show that
the Renyi entropies in this limit are determined by the area of extremal surface bounding the boundary region. This result implies that the von Neumann entropy of our tensor network agrees with the HRT formula of holographic theories (while the Renyi entropies generically do not agree\cite{dong2016gravity,dong2016deriving}). This
approach can also be generalized to include bulk quantum fields, and
compute bulk-boundary correlation functions. We show that the duality
defined by this setup satisfies the properties of the bidirectional
holographic code\cite{yang2015,hayden2016}, but has the advantage that
correlation functions can be studied for general space-time points,
rather than being restricted to a time slice.

The remainder of the paper is organized as follows. In
Sec.\ref{sec:GS}, we describe the general setup of the space-time
random tensor network. Starting from a bulk parent theory, we show how the boundary partition function and correlation functions are obtained by introducing random projections in the bulk. In Sec.\ref{sec:Renyi2}, we study the second Renyi entropy of a boundary region and show its relation to the partition function of a discrete gauge theory. We discuss how HRT formula is reproduced in proper large bond dimension limit. 
In Sec.\ref{sec:Renyin}, we generalize the discussion to $n$-th Renyi entropy. In Sec.\ref{sec:operators}, we
discuss the operator correspondence between bulk and boundary. We discuss how to define code subspace operators in the bulk and show the entanglement wedge reconstruction of such operators on the boundary. We also discuss the behavior of generic bulk-boundary and boundary-boundary correlation functions. As a special case of correlation structure, we also studied how the unitarity of the boundary theory depends on properties of the bulk theory.
In Sec.\ref{sec:fluctuations}, we discuss how to introduce proper gauge fixing in the definition of random tensor network, which is essential for bounding fluctuations and justify the semi-classical approximation. 
In Sec.\ref{sec:conclusion}, we summarize
this paper and discuss several open questions.

\section{General setup}\label{sec:GS}

\subsection{An overview of space-time tensor networks}


A tensor network is mathematically equivalent to a Feynman diagram. Each vertex that is adjacient to $k$ links is a rank-$k$ tensor $V^{i_1i_2...i_k}$ with each label $i_s$ defined on a link connecting to the vertex. When two neighboring vertices are connected by one link, the corresponding labels are contracted, leading to a new tensor
\begin{eqnarray}
V_x^{i_1i_2...i_p}V_y^{j_1j_2...j_{q}}\xrightarrow{\text{contraction}} g_{i_1j_1}V_x^{i_1i_2...i_p}V_y^{j_1j_2...j_{q}}
\end{eqnarray}
To define the contraction, one shall specify a metric $g_{ij}$ on each link. Without loosing generality, one can always assume the metric to be non-singular, since a singular metric can be viewed as a contraction in a lower dimensional subspace with a non-singular metric. Furthermore, one can always transform a non-singular metric to the standard form $g_{ij}=\delta_{ij}$ by a transformation on the vertex tensors. Therefore in the following we will always take the link metric to be the $\delta$-function, such that the information about the network is all encoded in the vertex tensors and the geometry of the network.


A tensor network in space-time is a
discrete version of path integrals, which can be used to define the
partition function of a statistical model, as is shown in
Fig. \ref{fig:TNofQFT}. (As examples of recent works on space-time tensor networks, see Ref.~\cite{levin2007tensor, gu2008tensor, jiang2008accurate, evenbly2015tensor}.) For simplicity we draw two-dimensional
networks, but all our discussions apply to general dimensions. For
concreteness, in Fig. \ref{fig:TNofQFT} we wrote the explicit
definition of the tensor which corresponds to the two-dimensional
Ising model. The degrees of freedom are defined on links of the graph,
which has dimension $D$. ($D=2$ for the Ising model.) Physical
properties correspond to multipoint correlation functions, which can
be computed by inserting operators (yellow boxes in
Fig. \ref{fig:TNofQFT} (a)) in links of the tensor networks. For example in
the Ising model if one wants to compute the multipoint spin
correlations, the operator to insert is the Pauli matrix $\sigma_z$,
which has the matrix element
$\left(\sigma_z\right)^{s_1s_2}=\delta_{s_1s_2}s_1$. A tensor network
can be defined on an arbitrary graph. For a generic curved space, one
can introduce a triangulation of the space and define a tensor network
on it, which is a discrete version of a quantum field theory on curved
space.

\addfig{5in}{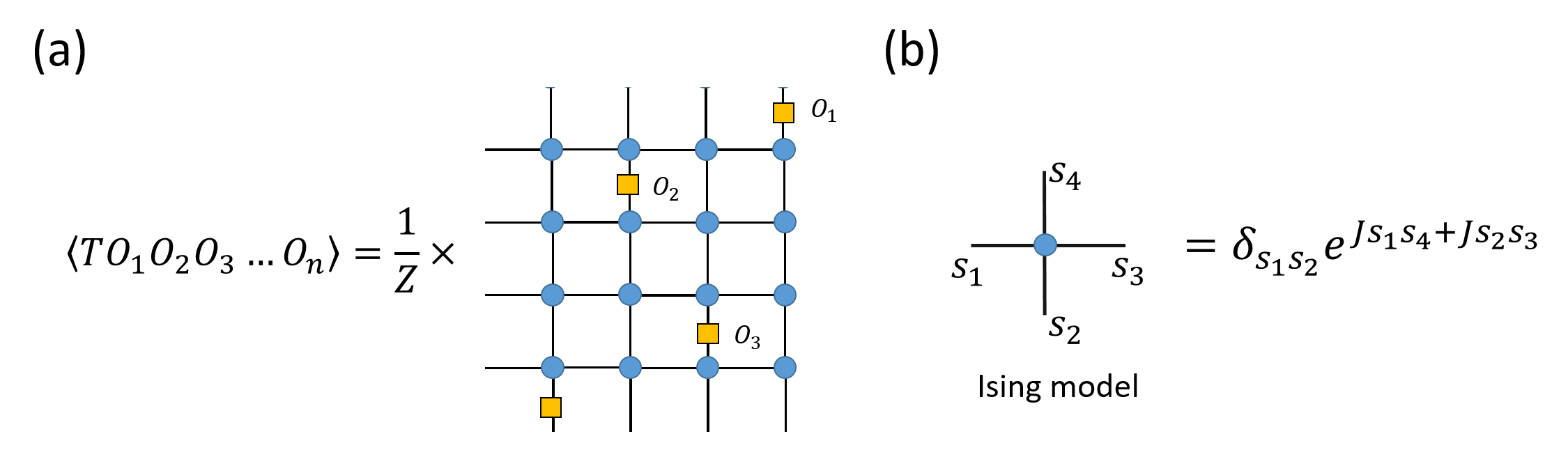}{(a) Illustration of a space-time tensor network which defines a statistical model, with the degrees of freedom defined on each link. With operator insertions $O_1,O_2,...,O_n$ (yellow boxes), evaluating this tensor network computes multipoint correlation functions in this system. (b) The definition of tensor that corresponds to the two-dimensional ferromagnetic Ising model. \label{fig:TNofQFT}}

\subsection{The holographic space-time tensor network}

Now we consider a different tensor network, which we propose to describe a holographic theory. We would like to define a tensor network on a $d+2$-dimensional space with a boundary, as is shown in Fig. \ref{fig:TNofgrav}. The main difference from the tensor networks in Fig. \ref{fig:TNofQFT} is a random projection $P_{xy}$ defined at each link $\overline{xy}$ in the bulk, as is shown by the red arrows.\footnote{As will be discussed in Sec. \ref{subsec:concept} and Sec. \ref{sec:fluctuations}, more precisely one should define the random projection on all links except those on a subgraph chosen by gauge fixing. However we decide to keep the simpler but imprecise definition here since it does not affect any result except for the discussion on fluctuations in Sec. \ref{sec:fluctuations}. } To be more precise, the random projection is an operator acting on the link Hilbert space with the form
\bea
P_{xy}=\ket{\Psi_{xy}}\bra{\Psi_{xy}}
\eea
with $\ket{\Psi_{xy}}$ a random state in the Hilbert space of the link. The tensor at the vertex is not random. At this moment we will leave the vertex tensor general, and discuss different choices of it later. 

The boundary links of the tensor network are not acted by the random projection. Instead, generic operators $O_1,O_2,...,O_n$ can be inserted to these links. Our proposal is to take the tensor network with boundary insertions $O_1,O_2,...,O_n$ and bulk random projections at each link as the definition of the multipoint function $\avg{TO_1O_2...O_n}_\pa$. Since the random projections can also be considered as operator insertions in the bulk theory, we can summarize this proposal as
\bea
\avg{TO_1O_2...O_n}_\pa=\frac{\avg{TO_1O_2...O_n\prod_{\overline{xy}}P_{xy}}_{\rm bulk}}{\avg{T\prod_{\overline{xy}}P_{xy}}_{\rm bulk}}\label{eq:dictionary}
\eea
where $\avg{...}_{\rm bulk}$ denotes the multipoint function in the bulk theory defined by the vertex tensors. 

Before studying the properties of this theory, we first provide some physical intuition why random projections in the bulk are a necessary ingredient. If we remove the random projections, Fig. \ref{fig:TNofgrav} defined multi-point correlation functions of an ordinary quantum many-body system in $d+2$ dimensions, with all operators supported on the boundary. It should be clarified that such a network clearly does not define a holographic duality. The discrepancy can be seen by considering equal time correlation functions, which are determined by the density matrix of the boundary. Since the boundary is simply a subsystem of the $d+2$-dimensional system, the boundary reduced density matrix is generically a mixed state, while a stand-alone $d+1$-dimensional system should have a pure state density matrix. The random projections are therefore the key feature that converts the tensor network in Fig. \ref{fig:TNofgrav} to a holographic theory, the properties of which will be studied in the rest of the draft. 

\addfig{4.5in}{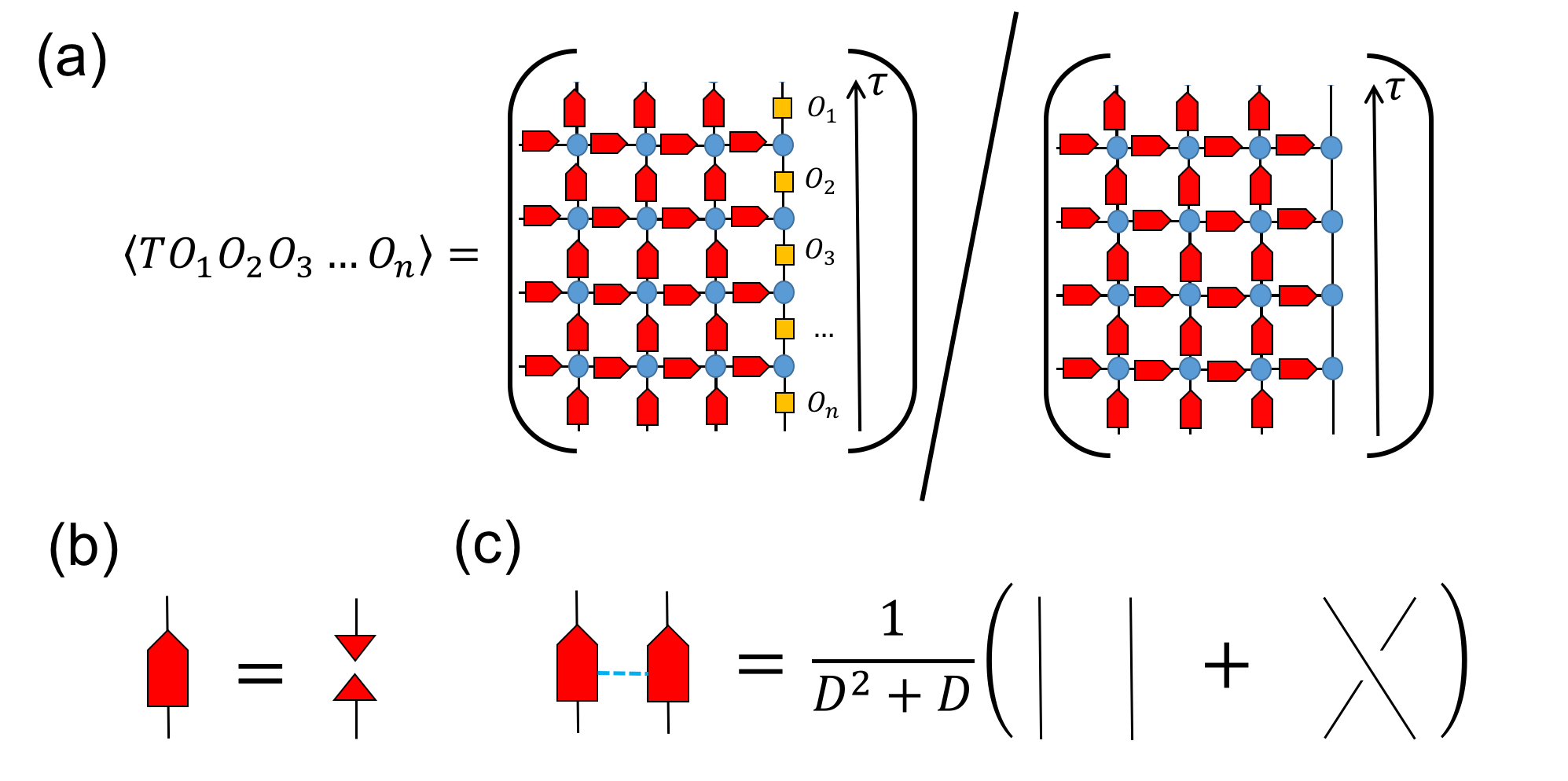}{\label{fig:TNofgrav} (a) The definition of multi-point functions of the boundary theory by a tensor network with one higher dimension. The red arrow along each link defines a random projector, as is illustrated in (b). (c) The random average over the direct product of two random projectors (denoted by the blue dashed line) results in the superposition of two operators, the identity operator and the swap operator, acting on the doubled Hilbert space. It should be noted that this figure has not taken into account the gauge redundancy subtlety discussed in Sec. \ref{subsec:concept} and Sec. \ref{sec:fluctuations}.} 

\subsection{Two conceptual subtleties}\label{subsec:concept}

Before studying the properties of the holographic tensor networks,
there are two conceptual points we need to discuss. First of all, the
random projections $P_{xy}$ on different links $xy$ of the space-time
network are all independent, so that for a given random realization,
the system after applying random projections does not have any
space-time symmetry. In particular, consider a bulk theory which has
time reflection symmetry before applying the random projections, so
that the forward time evolution from $t=-\infty$ to $t=0$ gives the
same state $\ket{\psi}$ as the backward time evolution from
$t=+\infty$ to $t=0$. For such a network, operator insertions at time
slice $t=0$ computes equal-time correlation functions, which are
determined by the density matrix $\rho=\ket{\psi}\bra{\psi}$. Now with
random projections that are independent at time $t$ and $-t$, the two
boundary states we obtained are generically different, which we denote
by $\ket{\psi_-}$ and $\ket{\psi_+}$, respectively. Now if we insert
an operator $O$ on the boundary, the network computes
$\bra{\psi_+}O\ket{\psi_-}={\rm tr}\kd{O\rho}$ with
\bea
\rho=\ket{\psi_-}\bra{\psi_+}\label{generalizedrho} 
\eea 
With these two states being different, $\rho$ is generically not a
density matrix since it is not Hermitian. Therefore our definition of
the boundary theory (\ref{eq:dictionary}) seems to be
problematic. However, similar to the case of spatial random tensor
networks\cite{hayden2016}, we are interested in certain limits where
the tensors have large bond-dimensions, and different random
configurations induce little fluctuation on entanglement
properties. For such large dimension tensor networks, we argue that
the space-time symmetry is restored in the sense of random average. This is
similar to how disorder breaks translation symmetry in an electronic
material, but all physical quantities that are self-averaged (such as
conductivity) appear translation invariant. With this understanding in
mind, in the following we will treat the boundary theory as a theory
with the space-time symmetry of the network before random projections.

Secondly, one apparent inconsistency in this theory is that, after the random projection the bulk vertices become disconnected with each other, so that their contribution to the partition function seems to be simply an overall numerical factor independent from the operator insertions. Therefore the bulk geometry seems to have trivial effect on the boundary theory, which is obviously not what we want. As we will see in next section, the random average of quantities such as Renyi entropy is mapped to a partition function of discrete gauge theory in the bulk, which does have a nontrivial dependence to the bulk geometry. Therefore the apparent inconsistency indicates that the physical quantities do not self-average, and there is a large fluctuation between different random configurations. As we will discuss in detail in Sec. \ref{sec:fluctuations}, this problem translates into a redundancy of the discrete gauge theory, which can be solved by a standard gauge fixing procedure. Translate back from the gauge theory to the random projections, the gauge fixing corresponds to removing the random projections on some bulk links which form a spanning tree of the bulk network. Since this gauge fixing only affects the fluctuation around large bond dimension saddle point, and it is much more natural to understand it after introducing the mapping to discrete gauge field, we opt to keep the imprecise definition of random projection on all links here, and ask the readers to keep in mind that actually the projections are applied to a subset of bulk links but the choice does not affect any result before Sec. \ref{sec:fluctuations}. 


\section{The second Renyi entropy calculation}
\label{sec:Renyi2}

\subsection{The second Renyi entropy of a boundary region}

\addfig{4.5in}{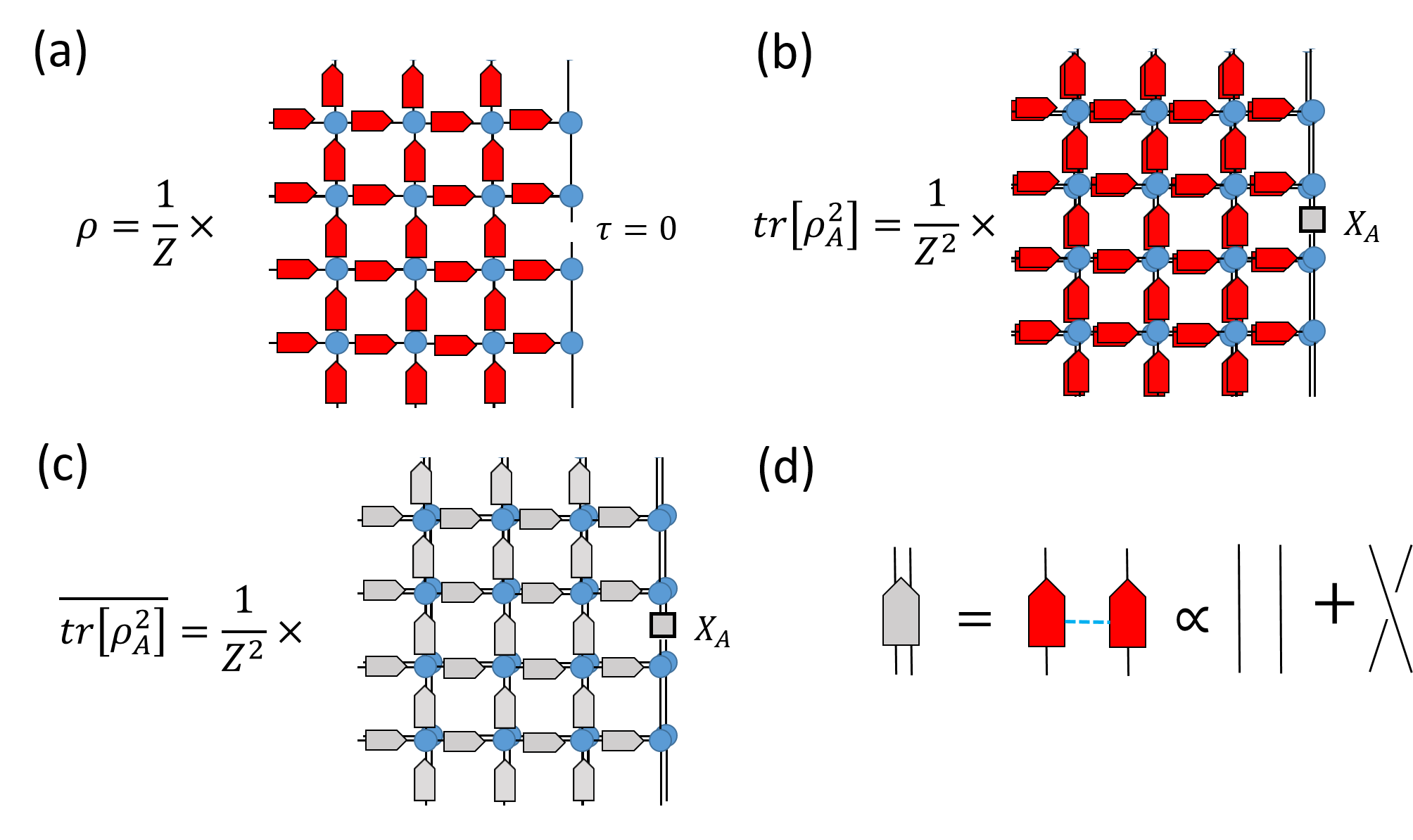}{\label{fig:Renyi2}(a) The network representation of the boundary density matrix $\rho$, with open legs at the $\tau=0$ cut representing the two indices of $\rho$. (b) For a boundary region $A$, ${\rm tr}\kd{\rho_A^2}={\rm tr}\kd{\kc{\rho\otimes \rho}X_A}$ (see text), which is represented by two copies of the same tensor network with an insertion of boundary operator $X_A$ (defined in Eq. (\ref{eqXA})). (c) Carrying the random average over the random projectors at each link following Fig. \ref{fig:TNofgrav} (c), one obtains the partition function of a $Z_2$ gauge theory. The grey arrow is defined in (d) as the average of $P_{xy}\otimes P_{xy}$. }

To relate the space-time tensor network to entanglement properties that we are familiar with, we consider equal time correlation functions at time $\tau=0$. Since in the computation of equal time correlation there is no insertions at other time steps, we can simply write $\avg{O}={\rm tr}\left(\rho O\right)$ for any operator, with the density matrix $\rho$ given by the network in Fig. \ref{fig:Renyi2} (a) with open boundary links in region $A$. \footnote{As is discussed in Sec. \ref{subsec:concept}, we are treating $\rho$ as the density matrix although in a particular random realization it may be non-Hermitian. In general, we can also consider space-time geometries without time reflection symmetry, so that $\rho$ is non-Hermitian even after the random average. All our results apply to such general situations.} 
We now compute the entanglement entropy of an arbitrary boundary region $A$. (Although we have been representing the boundary as a single line for simplicity, it should be remembered that the boundary is generically a $d+1$-dimensional system with $d$ spatial dimensions extending perpendicular to the paper plane.) The second Renyi entropy can be computed by
\bea
e^{-S_2(A)}&=&{\rm tr}\kd{\kc{\rho\otimes \rho}X_A}\label{eqXA}
\eea
with $X_A$ the partial swap operator that permutes the two copies of systems in $A$ region, but preserves the rest of the system \cite{hayden2016}. This expression of $e^{-S_2(A)}$ corresponds to Fig. \ref{fig:Renyi2} (b). Now if we take the random average over the link random projectors, we obtain Fig. \ref{fig:Renyi2} (c), in which each link projector $P_{xy}\otimes P_{xy}$ in the doubled system is replaced by its average value given in Fig. \ref{fig:TNofgrav} (c). Since the average is a sum over two operators, the whole tensor network can now be viewed as a partition function of Ising-like spin variables $\sigma_{xy}=\pm 1$ defined on each link, with $\sigma_{xy}=+1$ and $-1$ labeling the choice of identity channel and swap channel at this link, respectively. 

Interestingly, independent from the details of vertex tensors, the bulk statistical model of $\sigma_{xy}$ obtained by random averaging is always a {\it gauge theory} with $Z_2$ gauge invariance. The $Z_2$ gauge transformation at a given site $x$ is defined by changing $\sigma_{xy}\rightarrow -\sigma_{xy}$ for all neighboring links $\overline{xy}$, which is equivalent to interchanging the label of the two vertex tensors at $x$, which therefore preserves the partition function. The choice of boundary region $A$ defines a boundary condition of the $Z_2$ gauge vector potentials. Due to the gauge symmetry, the only gauge invariant information in this boundary condition is the location of $Z_2$ flux. Since $X_A$ is defined as the swap operator in region $A$ and identity elsewhere, the $Z_2$ flux at the boundary rests at the boundary $\partial A$, which is a co-dimension $2$ surface at the boundary. This situation is illustrated in Fig. \ref{fig:flux3d} for a system with bulk dimension $2+1$. 

\addfig{2.5in}{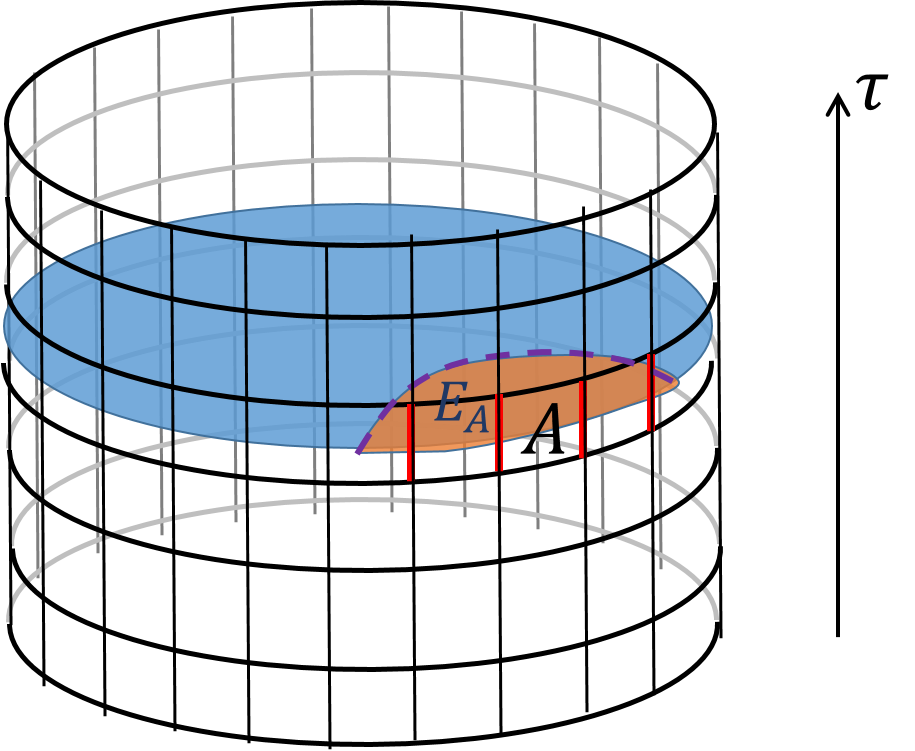}{\label{fig:flux3d} An illustration of the gauge field configuration $\ke{\sigma_{xy}}$. For a boundary region $A$ (the boundary of the organge region in the bulk), the boundary condition of $\sigma_{xy}$ is $\sigma_{xy}=-1$ for the red links across $A$, and $\sigma_{xy}=1$ elsewhere on the boundary. This boundary condition induces a flux co-dimension-$2$ surface $\gamma_A$ bounding the boundary of $A$, which is the purple dashed line. A gauge choice can be made by choosing $\sigma_{xy}=-1$ in the bulk for all links crossing a co-dimension $1$ surface $E_A$ which bounds $\gamma_A\cup A$, and $\sigma_{xy}=1$ elsewhere.}

With this boundary condition, the bulk partition function describes quantum fluctuations of the $Z_2$ gauge field, the action of which is determined by integrating out the bulk ``matter field" given by the vertex tensors. By adjusting the vertex tensors, one can obtain different dynamics of $Z_2$ gauge field. Before specifying the vertex tensor, one can already see some interesting property of the Renyi entropy. Let us assume that a bulk matter field was chosen such that the gauge field is in the weakly coupled limit. For example, this can be achieved by taking $N$ flavor of bulk fields (so that the vertex tensor is a direct product of $N$ tensors, each with a finite dimension) and considering the large $N$ limit. The effective action of $\sigma_{xy}$ will be $N\mathcal{A}_1\kc{\kd{\sigma_{xy}}}$ with $\mathcal{A}_1$ induced by a single flavor. Therefore in the large $N$ limit the gauge field is weakly coupled, and the electric flux induced by the boundary condition (the purple dashed line in Fig. \ref{fig:flux3d}) does not fluctuate. With this assumption, the electric flux will rest at the co-dimension $2$ surface $\gamma_A$ which minimizes the classical action of the bulk. Although the detail of this energy depends on the choice of bulk matter field, a general observation is that this classical action is actually the second Renyi entropy of a bulk region. To see that, one can take a gauge choice in the bulk by choosing a surface bounding $\gamma_A\cup A$, which we denote as $E_A$ in Fig. \ref{fig:flux3d}. $\sigma_{xy}$ can be chosen to be $-1$ for all links crossing this surface, and $+1$ everywhere else. In this choice, the bulk tensor network evaluates $e^{-S_2\kc{E_A}}={\rm tr}\kd{\rho_{E_A}^2}$, where $\rho_{E_A}={\rm tr}_{\overline{E_A}}\kd{\rho_b}$ is the reduced density matrix of bulk region $E_A$, and $\rho_b$ is the state of the bulk defined by contracting all tensors while leaving the indices open at a spatial surface (the disk with the blue and orange regions). For example, if we take the vertex tensors to be $N$ copies of the Ising tensors in Fig. \ref{fig:TNofQFT} (b), and take the network to be infinite and translation invariant along the imaginary time direction $\tau$, $\rho_b$ is the ground state of $N$ independent copies of the Ising model. 

In summary, without specifying the vertex tensor, we obtain the following general equation
\bea
S_2\kc{A}=S_2^{\rm bulk}\kc{E_A}
\eea
as long as the gauge field is in the weakly coupled limit. In the following we will pick an simplest choice of the vertex tensors, for which the action of $Z_2$ gauge field can be explicitly obtained, and the Renyi entropy satisfies Ryu-Takayanagi formula asymptotically.

\subsection{The bulk valence bond solid state and the HRT formula}\label{subsec:RT}

As a specific example of the general results in the previous section, we consider a very simple tensor network defined in Fig. \ref{fig:loop} (a). A closed loop is assigned to each $2$-dimensional plaquettes of the network, which can be viewed as the world line (in the Euclidean space-time) of a qudit with dimension $D$. \footnote{Obviously, to make this state well-defined, one needs to specify not only the vertices and links in the network, but also plaquettes.} If a link $xy$ is adjacient to $m$ plaquettes, there are $m$ loops passing through link $xy$, so that the dimension of the link $xy$ is taken to be $D^m$. The vertex tensor simply passes each qudit along the direction of the loop. One can denote all plaquettes adjacient to a vertex $x$ by $I=1,2,...,n$, and label the states of the two qudits at the two links which adjacient to both the plaquette $I$ and the vertex $x$ by $\mu_I=1,2,...,D$ and $\nu_I=1,2,...,D$, as is shown in Fig. \ref{fig:loop} (b). Then the explicit definition of the vertex tensor is
\bea
T_{\mu_1\nu_1\mu_2\nu_2...\mu_n\nu_n}=\delta_{\mu_1\nu_1}\delta_{\mu_2\nu_2}...\delta_{\mu_n\nu_n}
\eea

To understand the physical meaning of such a network, we can consider a network with translation symmetry in the imaginary time direction (such as the one in Fig. \ref{fig:loop} (a)), in which case the network defines an imaginary time evolution $e^{-\tau H}$. It should be reminded that we are now talking about the network without the random projections, which defines a bulk QFT. Denoting the time difference of two neighboring steps as $\tau_0$, we see that (up to normalization) $e^{-\tau_0H}$ is a projection operator to maximally entangled EPR pairs at each link:
\bea
e^{-\tau_0H}=\prod_{\overline{xy}}\ket{\overline{xy}}\bra{\overline{xy}},~\ket{\overline{xy}}=\frac1{\sqrt{D}}\sum_{\alpha=1}^D\ket{\alpha}_x\otimes \ket{\alpha}_y
\eea
Equivalently, one can write $H=U\sum_{\overline{xy}}\kc{1-\ket{\overline{xy}}\bra{\overline{xy}}}$ and take $U\rightarrow +\infty$. In this tensor network, at any spatial cut one obtains a state $\prod_{\overline{xy}}^\otimes\ket{\overline{xy}}$, which consists of a fixed configuration of EPR pairs. We follow the convention in the literature \cite{rokhsar1988superconductivity,ran2007projected,chen2012symmetry} and call such a state a valence bond solid (VBS) state. 

For the VBS state, the bulk action of the gauge field can be explicitly computed. When we consider the doubled theory and take the random average as is shown in Fig. \ref{fig:Renyi2} (c), each plaquette contributes a term
which is determined by $\prod_{\square}\sigma_{xy}$, i.e., the flux of the $Z_2$ gauge field in that plaquette. If 
$\prod_{\overline{xy}\in\square}\sigma_{xy}=+1 (-1)$, the contraction of loops around this plaquette gives the statistical weight $D^2$ ($D$), respectively. Therefore we obtain
\bea
e^{-S_2\kc{A}}&=&\frac1{Z}\sum_{\ke{\sigma_{xy}}}e^{-\mathcal{A}\kd{\ke{\sigma_{xy}}}}\nonumber\\
\mathcal{A}\kd{\ke{\sigma_{xy}}}&=&-\frac12\log D\sum_{I\in{\rm plaquettes}}\prod_{\overline{xy}\in I}\sigma_{xy}\label{eqRT}
\eea
which is the standard action of $Z_2$ gauge theory. One should be reminded that $\sigma_{xy}$ has the boundary condition set by the choice of boundary region, as was illustrated in Fig. \ref{fig:flux3d}. In the denominator, the partition function $Z$ has the same expression $\sum_{\ke{\sigma_{xy}}}e^{-\mathcal{A}\kd{\ke{\sigma_{xy}}}}$ but with the boundary condition $\sigma_{xy}=1$ everywhere on the boundary. Therefore Eq. (\ref{eqRT}) tells us that the Renyi entropy of a region $A$ is given by the action cost of adding an electric flux threading the boundary of $A$. 

\addfig{4.5in}{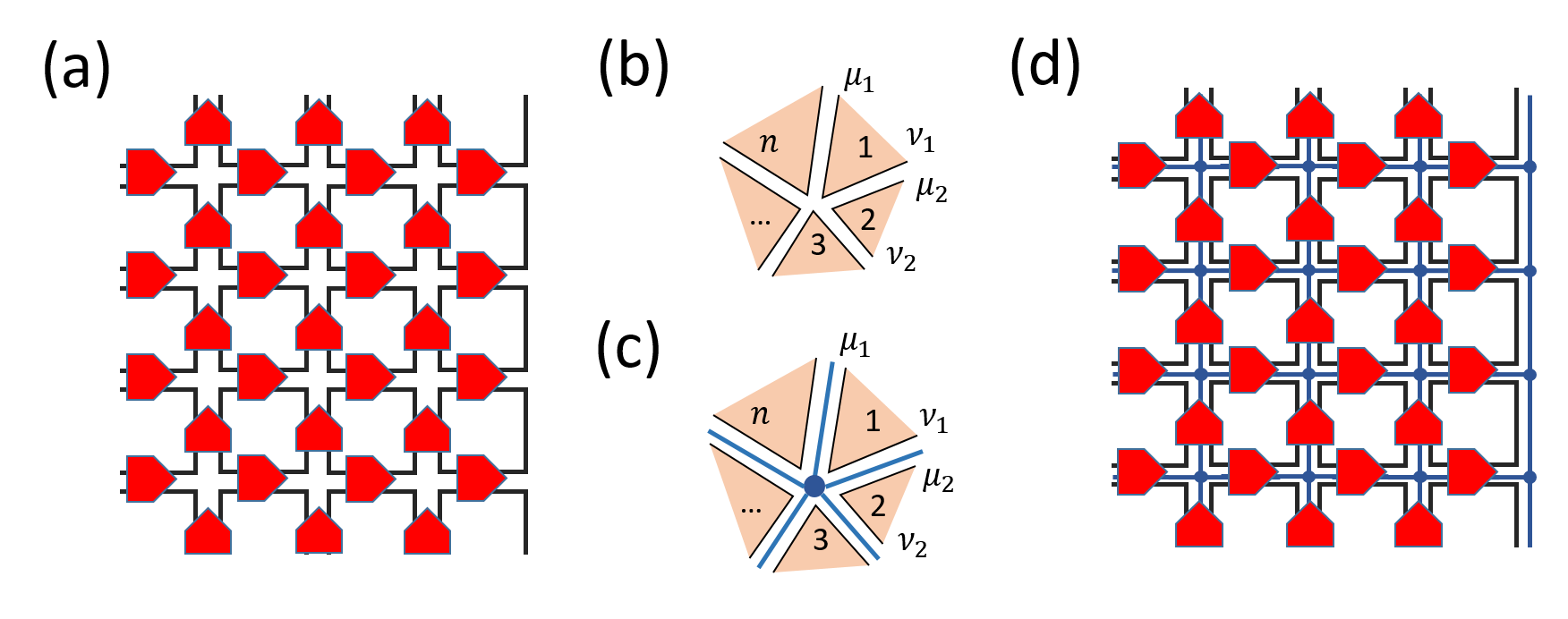}{(a) The tensor network that corresponds to the loop state we consider in Sec. \ref{subsec:RT}. The red arrows stand for random projectors, the same as in Fig. \ref{fig:TNofgrav}. (b) A more explicit definition of the vertex tensor, for a generic vertex that is adjacient to $n$ plaquettes. Each plaquette are adjacient to two links, which are labeled by $\mu_i,\nu_i$, $i=1,2,...,n$. $\mu_i,\nu_i=1,2,...,D$. (c) The generalization of loop state by introducing bulk entanglement, as is discussed in Sec. \ref{subsec:RTwithbulk}. Now each link is labeled by the loops passing through it, and an additional index $a=1,2,...,D_b$ labeling the remaining ``quantum field theory" degrees of freedom in the bulk. (d) An example of a tensor network with loops and additional bulk degrees of freedom.\label{fig:loop}}

In this action, $\log D$ plays the role of coupling constant, and the gauge field is weakly coupled in the large $D$ limit. In this limit, the electric flux induced by the boundary condition is heavy and classical. The classical action of an electric flux at a surface $\gamma_A$ is simply $\log D\abs{\gamma_A}$ with $\abs{\gamma_A}$ the area of $\gamma_A$. Therefore the lowest energy configuration is given by the minimal area surface (which we will also denote as $\gamma_A$), and the entropy is given by 
\bea
S_2\kc{A}\simeq \log D\abs{\gamma_A}
\eea
In other words, we have explicitly proved that the second Renyi entropy satisfies the covariant HRT formula (in Euclidean space-time) when the bulk is in the VBS state in the large $D$ limit. This result is the space-time analog of the RT formula in spatial random tensor networks discussed in Ref. \cite{hayden2016}, but the space-time approach is covariant, so that the entanglement entropy can now be computed for any boundary region $A$ in the space-time, rather than those restricted to a fixed spatial surface. If we consider a bulk which is a discretization of hyperbolic space $H^{d+2}$, the corresponding boundary theory will have the full conformal symmetry (in Euclidean signature) $SO(d+2,1)$ at scales much larger than the discretization, which was not possible in the spatial random tensor network approach. 

Furthermore, we would like to point out that even with Euclidean signature, where the HRT surface is a minimal surface, the space-time tensor network satisfying HRT is still not a trivial reformulation of the spatial tensor networks satisfying RT formula. For a generic geometry, different boundary regions can correspond to HRT surfaces that do not belong to a co-dimensional $1$ spatial slice of the bulk, which is by construction always true  for a spatial tensor network. The space-time formalism includes the spatial tensor network as a special case, when the bulk geometry has time reflection symmetry and all minimal surfaces bounding regions on the boundary time-reflection-symmetric slice lie in the bulk time-reflection-symmetric surface. 

\subsection{Some more comments on Lorenzian time}

For concreteness we have focused on Euclidean time, where the VBS state allows us to explicitly obtain the bulk gauge field action. However, physically our approach applies also to Lorenzian space-time geometry. The signature of the bulk theory is completely determined by the properties of vertex tensors. For a Lorenzian theory with unitarity, the vertex tensors should be unitary quantum gates. For example a unitary tensor network can be obtained by Trotter-Suzuki decomposition of a Hamiltonian time evolution\cite{trotter1959product,suzuki1976generalized,suzuki1976relationship}. After introducing the random projection on links and compute the random average of $\tr{\rho_A^2}$ for a boundary region, one still obtain a $Z_2$ gauge field coupled to bulk matter. The effective action of the gauge field is obtained by integrating out the bulk matter, except that the path integral ({\it i.e.} contraction of bulk tensors) is now carried in a Lorenzian theory. The Lorenzian analog of the VBS state is a short-range correlated theory, such as a boson with a large mass. For example if we consider $N$ flavors of massive bosons $\phi_i,~i=1,2,...,N$, the double-copied theory contains $\phi_i^{(1)},~\phi_i^{(2)}$ and the $Z_2$ gauge field couples to bosons by permuting the replica index. Integrating out the bosons, we expect to obtain a (discretized) Maxwell action for the $Z_2$ gauge field, which is weakly coupled in the large $N$ limit. Once the Maxwell action is obtained, the calculation of $\tr{\rho_A^2}$ still reduces to evaluating the effective action of an electric flux pinned to $\partial A$. With a Maxwell action in the weakly coupled limit, we expect
\bea
\tr{\rho_A^2}=\sum_{\gamma}e^{-\frac{N}{g}\abs{\gamma}}
\eea
where the sum is over all flux configurations $\gamma$ that terminates at $\partial A$, $\abs{\gamma}$ is the area of $\gamma$, and $g$ is some coupling constant. In the large $N$ limit, the sum will be dominated by the saddle point surface $\gamma_A$, although $\gamma_A$ is not a minimal surface any more. Therefore we expect the HRT formula to apply for general geometries as long as the bulk theory is $N$ independent copies of a short-range correlated theory, in the large $N$ limit. 

In the following discussion we will still use Euclidean signature and the VBS state as an explicit example, except in Sec. \ref{sec: unitarity}, but all discussions can be carried in parallel in Lorenzian time.

\subsection{Bulk entanglement corrections to the second Renyi entropy}\label{subsec:RTwithbulk}

In Ref. \cite{hayden2016}, it was shown that the random tensor networks can be used to define a holographic mapping, which a network with both bulk indices and boundary indices that maps a bulk state to a boundary state. This is more generic than a random tensor network that directly defines a state on the boundary. For a given holographic mapping network and a given bulk state $\rho_b$, the boundary region Renyi entropy $S_n$ in the large $D$ limit are shown to satisfy RT formula with bulk entanglement corrections. The bulk state contribution is a subleading term in $\log D$, but it is essential for many things, such as finite mutual information between two far-away boundary regions. Here we will show that the space-time tensor network approach can also take into account a more general bulk state, which gives similar corrections to the Renyi entropies. 

For this purpose, consider the definition of vertex tensor defined in Fig. \ref{fig:loop} (c) and (d). In addition to loops running around plaquettes, there is another tensor (the blue dot and blue lines in Fig. \ref{fig:loop} (c)) which describes some additional bulk degrees of freedom. Denote the dimension of each blue line as $D_b$, the dimension of each link is now $D^mD_b$ with $m$ the number of plaquettes adjacent to the link. Physically, if we take $D\rightarrow \infty$ and keep $D_b$ finite, the loops contribute a large amount of short-range entanglement, while the remaining degrees of freedom labeled by the blue line can be viewed as a ``bulk quantum fields", which made a smaller but possibly longer-ranged contribution to quantum entanglement between different bulk regions. In the following we will refer the degrees of freedom with dimension $D_b$ as the ``bulk quantum fields". 

After the random average in the second Renyi entropy calculation, the $Z_2$ gauge field $\sigma_{xy}$ is now coupled to both the loops and the bulk quantum fields. The action of the $Z_2$ gauge field is given by a sum of these two contributions:
\bea
\mathcal{A}\kd{\ke{\sigma_{xy}}}=-\frac12\log D\sum_{I\in{\rm plaquettes}}\prod_{\overline{xy}\in I}\sigma_{xy}+S_2^{L}\kd{\ke{\sigma_{xy}=-1}}\label{actionbulk}
\eea
Here $S_2^{L}\kd{\ke{\sigma_{xy}=-1}}$ denotes the second Renyi entropy of the co-dimension $1$ surface $\Sigma$ which crosses all links with $\sigma_{xy}=-1$, for the bulk quantum fields. \footnote{To be more precise, each link is dual to a co-dimensional $2$ surface in the dual graph, and $\Sigma$ is the union of them.} Consequently, in the large $D$ limit the gauge field is still weakly coupled, and the partition function is dominated by the flux configuration threading the minimal surface $\gamma_A$. In this limit, the last term simply gives the entanglement entropy of region $E_A$ in the state of bulk quantum fields, with $E_A$ is a Cauchy slice of the entanglement wedge of $A$, as is illustrated in Fig. \ref{fig:flux3d}. In summary, in this limit we obtain
\bea
S_2\kc{A}\simeq \log D\abs{\gamma_A}+S_2^L\kc{E_A}
\eea
which is in agreement with previous results\cite{faulkner2013quantum,hayden2016}. 

We would like to note that the formula also applies more generically if we take $D\rightarrow \infty$ and $D_b\rightarrow \infty$ simultaneously, except that now the two terms in action (\ref{actionbulk}) may compete with each other, and the minimal surface $\gamma_A$ is determined by minimizing the action, rather than the area.

\section{Higher Renyi entropies}
\label{sec:Renyin}

The generalization of the discussion above to higher Renyi entropies is straightforward. For the calculation of $n$-th Renyi entropy of a boundary region $A$ one would like to calculate 
\bea
e^{-(n-1)S_n\kc{A}}={\rm tr}\kd{\rho_A^n}={\rm tr}\kd{\kc{\rho_\partial^{\otimes n}}X_{nA}}
\eea
with $X_{nA}$ the cyclic permutation of the $n$ copies of systems in $A$ region. For the tensor network defined in Fig. \ref{fig:TNofgrav}, this calculation corresponds to taking $n$ copies of the network and insert an operator $X_{nA}$ at the boundary, in the same way how $X_A$ is inserted in Fig. \ref{fig:Renyi2} (c). 
The random average of this tensor network is determined by that of $n$ copies of random projectors:
\bea
\overline{P_{xy}^{\otimes n}}=\frac1{C_{n,xy}}\sum_{g_{xy}\in S^n}g_{xy}\label{nthmoment}
\eea
with the normalization constant $C_{n,xy}=\frac{\kc{D_{xy}+n-1}!}{\kc{D_{xy}-1}!}$, and $g_{xy}$ are the elements of permutation group with $n$ objects, which act on the $n$-copied system by permuting the indices of the $n$ copies. 

Following the same derivation as the second Renyi entropy case, this random average is thus mapped to the partition function of a $S^n$ gauge theory. The gauge invariance comes from the symmetry of permuting the vertex tensor at any vertex $x$. By choosing different bulk vertex tensors, one can obtain different dynamics of the $S^n$ gauge field. 

If we choose the same VBS state that was studied in Sec. \ref{subsec:RT}, the action of the gauge field has the standard form
\bea
\mathcal{A}\kd{\ke{g_{xy}}}=-\log D\sum_{I\in{\rm plaquettes}}\chi\kc{\prod_{\overline{xy}\in I}g_{xy}}\label{Snaction}
\eea
Note that the permutation group elements $g_{xy}$ is defined for an oriented link, with $g_{yx}=g_{xy}^{-1}$. The flux defined by the product $\prod_{\overline{xy}\in I}g_{xy}$ should have $\overline{xy}$ oriented properly. (For example one can choose $xy$ to be oriented such that the plaquette $I$ is always on the left of the link when one goes from $x$ to $y$. ) The function $\chi\kc{g}$ denotes the number of loops in a permutation element $g$. One can also write $\chi(g)=\frac{\log{\rm tr}\kc{g}}{\log D}$ with the trace taken in the same representation of $g$ as in Eq. \ref{nthmoment}.

Similar to the $n=2$ case, in the large $D$ limit, the non-Abelian discrete gauge theory (\ref{Snaction}) is in the weakly coupled limit, with the electric flux classical. The minimal action configuration consists of an electric flux threading the minimal surface $\gamma_A$, with the flux the cyclic permutation $C_n$. Therefore we obtain the HRT formula
\bea
S_n\kc{A}=\log D\abs{\gamma_A}
\eea

The generalization to systems with bulk entanglement, given by networks in Fig. \ref{fig:loop} (c) and (d), also directly applies to the $n$-th Renyi entropy. In the limit $D\rightarrow \infty$ and finite $D_b$, the $n$-th Renyi entropy of a boundary region is given by 
\bea
S_n\kc{A}=\log D\abs{\gamma_A}+S_n^{L}\kc{E_A}
\label{nthrenyi}
\eea

The discussion of large $D$ limit here assumes that a single
configuration (modulo gauge transformations that will be discussed in
Sec. \ref{sec:fluctuations}) dominates in the large $D$ limit. For a
geometry with a unique minimal surface, this assumption should be
correct. However, upon increasing $n$, the number of low energy
configurations increase quickly
Therefore, the large $D$ limit and large $n$ limit might be
non-commuting. We also would like to note that the $n$-independent
Renyi entropy in large $D$ limit is a feature shared by spatial random
tensor network states\cite{hayden2016}, but different from large $N$
gauge theories with gravitational dual\cite{dong2016gravity}. This
difference is related to the absence of gapless gravitons in the bulk,
and is a key ingredient of AdS/CFT correspondence that is missing in
the tensor network models.

\section{Operator correspondence between bulk and boundary}
\label{sec:operators}

\subsection{Bulk-boundary correlation functions}
\label{subsec:bulkboundary}

In the discussion so far we have focused on entanglement properties of the boundary state. Since different bulk states can be mapped to different boundary states, we can also discuss the mapping between bulk operators and boundary operators. We consider the network in Fig. \ref{fig:loop} (d), with bulk states consisting of loops with dimension $D$ and bulk quantum fields with dimension $D_b$. Instead of considering a boundary correlation function, we can consider a boundary-bulk correlation function, with operator insertions $O_a$ the boundary and $\phi_x$ in the bulk, as is shown in Fig. \ref{fig:errorcorrection} (a). It should be noticed that we are inserting operators in the bulk which acts only on the ``bulk quantum field" index. Physically one can think such operators as ``low energy operators" in the bulk. If we consider the effect of bulk operators $\phi_x$ as a perturbation to the bulk state, this boundary-bulk correlation function is computing the effect of such perturbations to a generic boundary correlation function. Since all such perturbations can be viewed as modifying the boundary theory, one expects that each multipoint operator in the bulk $\prod_x\phi_x$ corresponds to some boundary operator. 

\addfig{4in}{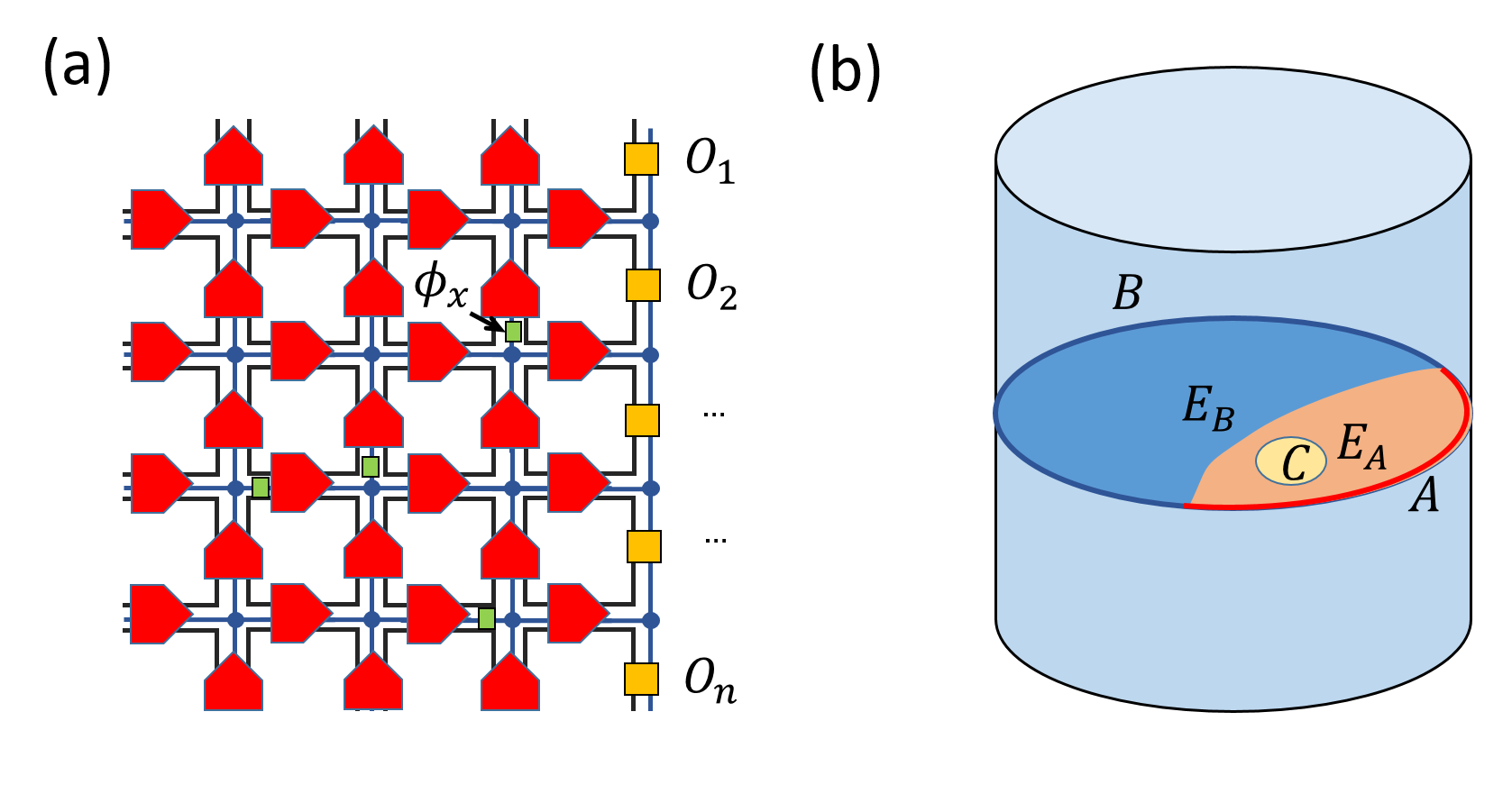}{\label{fig:errorcorrection} (a) The
  definition of bulk-boundary correlation function, with insertion of
  operators at the boundary (orange boxes) and bulk quantum field
  theory states (green boxes). (b) The choice of bulk region $C$
  (yellow disk) and boundary regions $A$ (red), $B$ (blue), all of
  which lies in a spatial surface $\Sigma$ (the whole disk). $E_A$ and
  $E_B$ are spatial slices of the entanglement wedge of $A$ and $B$, such that their
  interface is the minimal surface bounding $A$. }

\subsection{Equal time correlations and the error correction property}
\label{subsec:errorcorrection}

The reconstruction of the bulk operators, which lie in the
entanglement wedge of some boundary subregions, onto the corresponding
boundary regions is interpreted as a quantum error correction
code\cite{almheiri2014bulk,dong2016reconstruction,harlow2016ryu,dong2016bulk}
which explicitly illustrates the ``subregion-subregion duality" in the
context of AdS-CFT.  Roughly speaking, for a boundary region $A$,
there exists a code subspace $\mathcal{H}_{code}$ of the bulk Hilbert
space, and a subalgebra of operators acting on this subspace. For any
operator $O_a$ in this subalgebra, and any state
$\ket{\phi}\in\mathcal{H}_{code}$, there exists an operator $O_A$
acting on boundary region $A$ such that \bea M O_a\ket{\phi} = O_A
M\ket{\phi}, \eea
where $M$ is the holographic mapping from the bulk Hilbert space to the boundary Hilbert space.  Physically, the code subspace is the Hilbert space of the
effective field theory living in the boundary's entanglement
wedge. Explicit examples are shown in \cite{pastawski2015,yang2015,hayden2016,qi2017holographic}.

In the space-time tensor formalism, we define the code space
$\mathcal{H}_{code}$ as the space of all bulk states sharing the same background geometry, in the limit $D_b\ll D$. Thus, the states in the $\mathcal{H}_{code}$ share the same
entanglement wedge. In the following part, we will show that, within
$\mathcal{H}_{code}$, the operators that act on the entanglement
wedge of some boundary region $A$ can be reconstructed on $A$.

In the previous sections, we treat the VBS states as the backbone of
the bulk geometry, on top of which the  bulk quantum fields
live. However, from another perspective, we can treat both the VBS
states and the bulk quantum fields together as the bulk states. In
contrast with $S_n^{L}(E_A)$ in Eq.\ref{nthrenyi}, which denotes the
$n$th Renyi entropy of only the bulk quantum fields, we use
$S_n^{bulk}(E_A)$ to represent the entropy of all the bulk states in $E_A$. Thus in the $D\rightarrow \infty$ limit,
Eq.\ref{nthrenyi} can be rewritten as $S_n(A)=S_n^{bulk}(E_A)$. In
this section, all notations with super-index $bulk$ refers to the direct product of the bulk quantum fields and the VBS states.

In order to prove that the space-time tensor formalism have the error
correction property, one needs to prove that all operators within the code subspace, acting in the entanglement wedge of $A$, can be reconstructed on the
boundary $A$. According to Ref. \cite{dong2016reconstruction,harlow2016ryu,dong2016bulk}, we only
need to show that the relative entropy in the compliment of region $A$
and $E_A$ (denoted by $B$ and $E_B$, see Fig.\ref{fig:errorcorrection}
(b)) satisfies
$S(\rho_{B}|\sigma_{B}) =
S(\rho^{bulk}_{E_{B}}|\sigma^{bulk}_{E_{B}})$, where $\rho^{bulk}$ and
$\sigma^{bulk}$ are the density matrices of two states
$\ket{\rho^{bulk}}, \ket{\sigma^{bulk}}\in H_{code}$, and 
$\rho_{E_{B}}^{bulk} = \tr_{E_A}{(\rho^{bulk})}$.
$\rho_{B} =\tr_{A}(\rho)$, where $\rho$ is the boundary state
corresponding to the bulk state $\rho^{bulk}$ . Because
$S(\rho_B|\sigma_B) = - S(\rho_B) -
\tr_{B}\left(\sigma_B\log\rho_B\right)$, and
$S(\rho_B) = S(\rho_{E_B}^{bulk})$ (Eq. (\ref{nthrenyi})), we only need
to show that
$\tr_{B}\left(\sigma_B\log\rho_B\right)=
\tr_{E_B}\left(\sigma_{E_B}^{bulk}\log\rho_{E_B}^{bulk}\right)$. To study this quantity we introduce a replica trick
\bea
\tr_B\kc{\sigma_B\log\rho_B}=\lim_{n\rightarrow 1}\frac1{1-n}\tr\kc{\sigma_B\rho_B^{n-1}}
\eea
Therefore if $\tr{\sigma_B\rho_B^{n-1}}=\tr{\sigma_{E_B}\rho_{E_B}^{n-1}}$ for all $n$, by analytic continuation we have proved the relative entropy equality. 
\begin{figure}[ht!]
  \centering
  \includegraphics[width=0.8\textwidth]{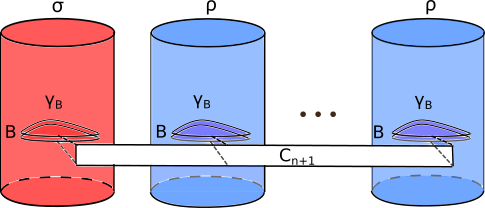}
  \caption{An illustration of
    $\overline{\tr_B{\left(\sigma_B\rho_B^{n}\right)}}$ in the
    space-time tensor network. The red and blue cylinders represent
    $\sigma$ and $\rho$, respectively, which share the same geometry
    and only differ in the bulk quantum field. The boundary condition
    is set by inserting a cyclic permutation operator $C_{n+1}$ in the
    boundary region $B$. In the $D_b\ll D$ limit, the leading
    contribution is from the VBS states. Thus the dominating classical
    flux configuration contains a flux following the minimal surface
    $\gamma_B$, and the gauge field variables on the links, modulo
    gauge transformation, are fixed as $C_{n+1}$ for links crossing a
    codimension-$1$ surface $E_B$ that bounds $B$ and $\gamma_B$, and
    identity elsewhere.  }
  \label{fig:relativeentropy}
\end{figure}

After the random average, we obtain
\begin{eqnarray}
 \nonumber
 &&  \overline{\tr_B\left(\sigma_B\rho_B^{n}\right)} = \frac{1}{\tr{\sigma}\left(\tr{\rho}\right)^n} \sum_{\{g_{xy}\}}e^{-A[\{g_{xy}\}]}\\
&&A[\{g_{xy}\}]= -\log D \sum_{I\in plaquettes} \chi\left(\prod_{\overline{xy}\in I} g_{xy}\right) - \log\left(\tr{\left(\sigma^L\left(\otimes\rho^L\right)^{n} \cdot \prod_{\overline{xy}\in I} g_{xy}\right) }\right)
\end{eqnarray}
The first term is the same as Eq. (\ref{Snaction}), and the second term
shows the contribution from the bulk quantum fields. In the $D\gg D_b$
limit, the first term dominates. Minimizing the first term leads to the classical configuration that contains the flux of cyclic permutation $C_n$ winding around the
co-dimensional two minimal surface (see Fig.\ref{fig:relativeentropy}). 
For such a spin configuration, we see that 
\begin{equation}
  \label{eq:sigmarho}
   \overline{\tr_B\left(\sigma_B\rho_B^{n}\right)} \simeq D^{-n|\gamma_B|}\cdot \tr_{E_B}\left(\sigma^L_{E_B}\left(\rho^L_{E_B}\right)^{n}\right)= \tr_{E_B}\left(\sigma^{bulk}_{E_B}\left(\rho^{bulk}_{E_B}\right)^{n}\right) ~~ \forall n\geq 0
\end{equation}
Thus we conclude
\begin{eqnarray}
  \tr_{B}\left(\sigma_B\log\rho_B\right)=
\tr_{E_B}\left(\sigma_{E_B}^{bulk}\log\rho_{E_B}^{bulk}\right) ~~
\Rightarrow ~~ S(\rho_B|\sigma_B) = S(\rho^{bulk}_{E_B}|\sigma^{bulk}_{E_B})
\label{eq:relativeentropy}
\end{eqnarray}
Details that derives Eq. (\ref{eq:relativeentropy}) from Eq. (\ref{eq:sigmarho}) is shown in Appendix.\ref{app:re}.

It should be noted that the terms on the
r.h.s of the Eq.\ref{eq:relativeentropy} all have the super-index
$bulk$, which include both the loop states and the bulk quantum fields
states. Therefore the error correction properties apply to not only the operators in the bulk quantum field theory, but also to more general bulk operators that can act on the entire bulk Hilbert space, as long as it does not change the entanglement too much to change the location of the minimal surface. For example, one can consider an operator acting on a short-range EPR pair across the minimal surface $\gamma_A$, which reduces its entanglement entropy by some amount $\Delta S\ll \log D$. If the minimal surface is unique, the location of minimal surface remains the same after this change, so that such an operator can be reconstructed from the boundary region $A$. In general, the subalgebra of operators that can be reconstructed on $A$ does not have a tensor factorization in real space. 


As a side remark, we would like to mention that the above proof
also directly applies to the spatial random tensor networks in Ref. \cite{hayden2016}.

\subsection{More general correlators}

The random average technique allows us to understand more refined structure of correlation functions. For this purpose we generalize the concept of correlation matrix defined in Ref. \cite{hayden2016}. For two space-like regions $A$ and $B$ (which does not need to be defined on the same time slice), each region defines a sub-algebra of operators. We denote $\{O_A^\alpha\} (\{O_B^\beta\})$ as a orthonormal basis of Hermitian operators in $A(B)$, which satisfies
\begin{equation}
  \label{eq:7}
  \tr[O_A^\alpha O_A^\beta]=\delta^{\alpha\beta},~~ \sum_{\alpha} [O_A^\alpha]_{ab}[O_A^\beta]_{cd}=\delta_{ad}\delta_{bc}
\end{equation}
The correlation matrix between the two regions is defined as
\begin{equation}
  \label{eq:5}
  M^{\alpha\beta} = \frac{\tr[\rho O_A^\alpha O_B^\beta]}{\tr[\rho]}
\end{equation}
Any two-point correlation function $\langle F_A G_B\rangle$ can be expressed as a bilinar form$\langle  F_A F_B \rangle = \sum_{\alpha \beta} f_{A\alpha} M^{\alpha \beta} g_{B\beta} $, where $f_{A\alpha}$ is the coefficient of $F_A$ expanded onto the basis $\{O_A^\alpha\}$, and similar for $g_{B\beta}$. Therefore $M^{\alpha\beta}$ contains complete information about correlation functions between $A$ and $B$. To define a local-unitary invariant measure of the correlation, we study 
\begin{equation}
  \label{eq:6}
  C_{2n} = \tr\left[\left(M^\dagger M\right)^n\right]
\end{equation}
$C_{2n}$ is independent of the choice of operator basis
$\{O_A^\alpha \}$, $\{O_B^\beta \}$. $C_{2n}$ for all $n$ determines all singular values of correlation matrix $M$. The singular values of $M$ corresponds to a special basis of operators $F_{An},~G_{Bm}$ satisfying $\langle F_{An}G_{Bm}\rangle=\delta_{nm}\lambda_n$, so that this basis is the analog of primary fields in conformal field theories, and the singular values $\lambda_n$ (which decays in power law in the CFT case) provide a complete measure of the correlation structure between $A$ and $B$. 

Using the orthonormal condition Eq. \ref{eq:7}, one obtains
\begin{equation}
  \label{eq:8}
  C_{2n} = \frac{\tr\left[\rho^{\otimes 2n}\left(\mathcal{X}_A\otimes\mathcal{Y}_B \right)\right] }{\left(\tr \rho\right)^{2n}}
\end{equation}
where $\mathcal{X}_A$ and $\mathcal{Y}_B$ are two permutation operators
\begin{equation}
  \label{eq:9}
  \mathcal{X}_A = (1~2)(3~4)\cdots(2n-1~2n),~ \mathcal{Y}_B = (2~3)(4~5)\cdots(2n~1)
\end{equation}
Therefore after random average, the calculation of $C_{2n}$ is mapped to a partition function of $S_{2n}$ discrete gauge theory in the same way as $2n$-th Renyi entropy, except for a different boundary condition. The boundary condition now consists of $\mathcal{X}_A$ flux ending at $\partial A$ and $\mathcal{Y}_B$ flux ending at $\partial B$, instead of a flux of cyclic permutation $(123...2n)$ for both $\partial A$ and $\partial B$ in the case of Renyi entropy calculation. In the large bond dimension limit, if the two regions are far away enough such that the minimal surface bounding $A\cup B$ is disjoint ({\it i.e.} when the mutual information $I(A:B)\simeq 0$), the flux surfaces for $C_{2n}$ will also be disjoint, consisting of a $\mathcal{X}_A$ ($\mathcal{Y}_B$) surface on minimal surface $\gamma_A$ ($\gamma_B$), respectively. This leads to 
\bea
C_{2n}=D^{-n\abs{\gamma_B}-n\abs{\gamma_A}}C_{2n}^{\rm bulk}(E_A,E_B)
\eea
where $C_{2n}^{\rm bulk}(E_A,E_B)$ is the same quantity defined for the bulk QFT between the entanglement wedge regions $E_A$ and $E_B$. If the bulk low energy degree of freedom is trivial, $C_{2n}^{\rm bulk}(E_A,E_B)=1$, in which case $C_{2n}$ corresponds to a flat spectrum of singular values at $\lambda_0=D^{-\abs{\gamma_A}/2}D^{-\abs{\gamma_B}/2}$. The factorized form suggests that there is no connected correlation functions. When there is a nontrivial $C_{2n}^{\rm bulk}(E_A,E_B)$, it contributes all nontrivial connected correlation functions. In other words, the correlation spectrum (singular value spectrum of two-point functions) of regions $A,B$ is identical to that of the bulk QFT in the large $D$ limit, up to an overall factor $D^{-\abs{\gamma_A}/2}D^{-\abs{\gamma_B}/2}$. It should be noted that the discussion here is independent from whether $A$ and $B$ are space-like separated, and only requires that the HRT surface bounding $A\cup B$ (which in our setup is well-defined even if $A$ and $B$ are not space-like separated) is disjoint. 

The above discussion can also be generalized from boundary-boundary
correlation functions to more general bulk-boundary correlation
functions.  For a given set of boundary points and bulk points, one
can always define a ``generalized density martrix" $\rho$ by
contracting all other indices in the tensor network and leave these
boundary points and bulk points open. For the network in
Fig. \ref{fig:errorcorrection} (a), this means to remove all boundary
operators and the bulk operators, leaving the corresponding indices
open. This ``generalized density matrix" determines all multi-point
correlations between these bulk and boundary points, in the same way
how an ordinary density matrix determines equal time
correlators. Specifically, we focus on the correlation function
between two regions and calculate the $C_{2n}$, for all $n$. 


\addfig{2.1in}{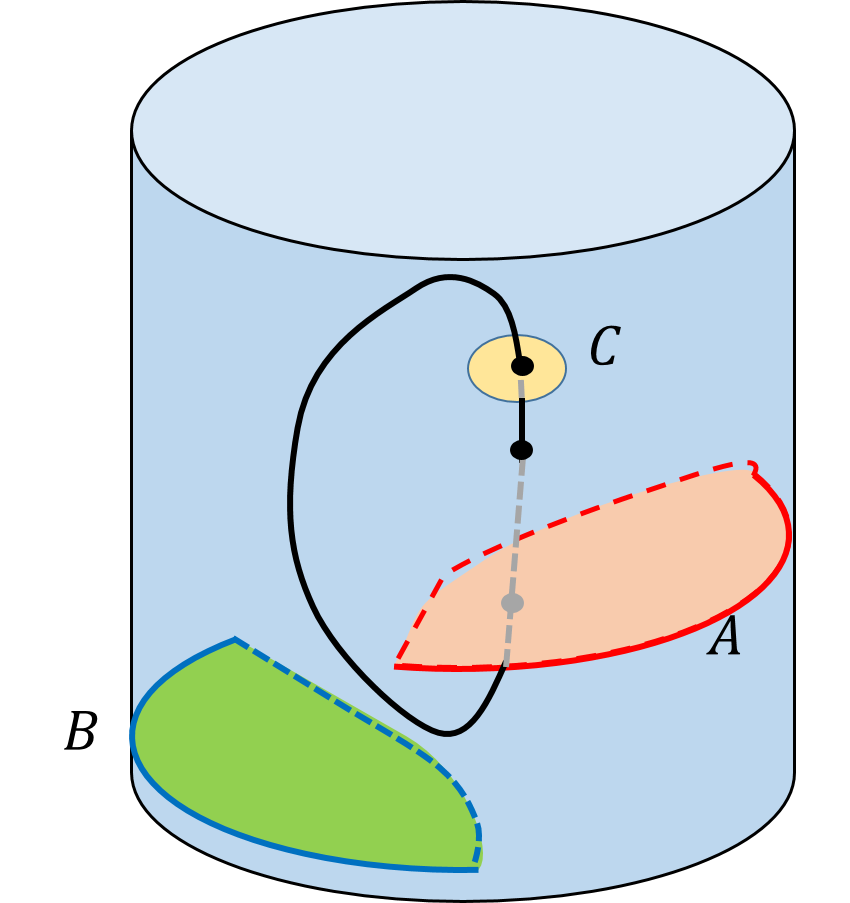}{An example of the bulk-boundary correlation between bulk region $C$ and boundary regions $A$ and $B$. The bulk theory contains a closed world line of a qudit that intersects with $C$ and the entanglement edge of $A$ but not with that of $B$ (see text). \label{fig:generalcorr}}
  
As a concrete example, we consider a very simple network, illustrated
in Fig. \ref{fig:generalcorr}. The bulk state is defined as
the direct product of a VBS state and a single pair of qudits with
dimension $D_b$. The qudits have a world line $l$ which forms a close
loop. $C$ is a bulk small region that intersects with the world line $l$. Two boundary regions $A$ and $B$ are chosen such that $l$ links with the closed co-dimension $2$ surface $\gamma_A\cup A$, but does not link with $\gamma_B\cup B$.  For such a state, we analyze the correlation spectrum between $B$ and $C$ and that between $A$ and $C$.

When we calculate $C_{2n}$ between $B$ and $C$, because the $\gamma_B$ doesn't link with the world line $l$, in the large $D$ limit,
\begin{equation}
  \label{eq:10}
  C_{2n}(B;C) = D^{-n|\gamma_B|} D_b^{-n}
\end{equation}
From this result, we know there is only one non-zero singular value of
matrix $M^{\alpha,\beta}_{B,C}$ in the large $D$ limit,
$\lambda = D^{-|\gamma_B|/2} D_b^{-1/2}$. One can compare this result with the disconnected piece of the correlation function:
\begin{equation}
  \label{eq:12}
  M^{\alpha \beta}_{dis} (B;C) =\frac{\tr[\rho O_B^\alpha]\tr[\rho O_C^\beta]}{\tr[\rho ]^2}
\end{equation}
which is a rank 1 matrix. We can define
\begin{equation}
  \label{eq:13}
  C_{2n,dis}(B;C) = \tr\left[\left(M_{dis}^\dagger M_{dis}\right)^n\right] = \frac{\left(\tr[\rho^{\otimes 2}\mathcal{X}_B] \tr[\rho^{\otimes 2}\mathcal{X}_C]\right)^n}{\tr[\rho ]^{4n}}
\end{equation}
where $\mathcal{X}_B$ is the swap operator acting in the region $B$. After random average, this quantity corresponds to a $S_{4n}$ gauge theory with the boundary condition of $\mathcal{X}_A$ acting on the first $2n$ replica and $\mathcal{Y}_B$ acting on the second $2n$ replica.  A straightforward calculation shows that in the large $D$ limit
\begin{equation}
  \label{eq:14}
  C_{2n,dis}(B;C) = D^{-n|\gamma_B|} D_b^{-n} = C_{2n}(B;C) 
\end{equation}
Thus we conclude that there is no connected correlation function between boundary region $B$ and the bulk region $C$ in the large $D$ limit.

The situation is totally different for the correlation function between boundary region $A$ and bulk region $C$. Similarly, in the large $D$ limit, 
\begin{equation}
  \label{eq:10}
  C_{2n}(A;C) = D^{-n|\gamma_A|} D_b^{1-2n}
\end{equation}
while the disconnected part is
\begin{equation}
  \label{eq:14}
  C_{2n,dis}(A;C) = D^{-n|\gamma_A|} D_b^{-2n} < C_{2n}(A;C)
\end{equation}
Thus there is non-trivial connected correlation between $A$ and $C$. Physically, the non-trivial correlation is a consequence of the fact that operators acting on the qubit with world line $l$ can be locally reconstructed on boundary region $A$.

\subsection{Comments on unitarity of the boundary theory}\label{sec: unitarity}


As we discussed earlier, our formalism applies to both Euclidean and Lorenzian space-time. In Lorenzian case an important question is whether the boundary theory defined by our tensor newtork is unitary. This question is closely related to the correlation function we discussed in previous subsections. In this section, we will show that boundary unitarity emerges when the bulk has enough degrees of freedom. In contrast, in the limit of small number of bulk degrees of freedom (which is the case if we restrict to the code subspace around a given classical geometry), the corresponding boundary theory only have unitarity in the code subspace. 



In order to decide whether the evolution is unitary, we treat the
evolution operator $\hat{U} = \sum_{i,j} U_{ij} \ket{i}\bra{j}$ as a
state in doubled system $\ket{U} \equiv \sum_{i,j} U_{ij} \ket{i}_{F}\otimes\ket{j}_{P}$ (Fig. \ref{fig:unitarity} (a))
and measure the mutual information $I(P:F)$ between the past subsystem (P) and the future subsystem (F). This mutual information is also the quantum channel capacity of $U$\cite{lloyd1997capacity}. If and only if 
the past state is maximally entangled with the future state, the
evolution operator $\hat{U}$ is unitary. To study the entanglement in $\ket{U}$, we first study the second Renyi entropy. It is obvious that 
\begin{equation}
  \label{eq:1}
  S_{F\cup P}^{(2)} =0
\end{equation}
since $\ket{U}$ is a pure state. Thus we only need to calculate the entropy of the past or the future state by inserting swap operators to the past or the future boundary. The corresponding bulk geometry is a bulk Lorenzian time evolution and its complex conjugate, connected at an initial time slice and a final time slice, as is shown in Fig.\ref{fig:unitarity}(a). (One can glue the forward and backward time evolution at an arbitrary bulk Cauchy surface that bounds the boundary initial time slice. Different choices of the Cauchy slice gives the same partition function due to unitarity of the bulk quantum field theory.) For concreteness we consider the entropy of the future system. This calculation is similar to
Sec.\ref{sec:Renyi2} except that the boundary condition now is $\sigma_{xy}=-1$ for all vertical links crossing the future boundary, which we denote as $F$, and $\sigma_{xy}=1$ elsewhere on the boundary (Fig. \ref{fig:unitarity} (b) and (c)). Therefore the only nontrivial $Z_2$ flux is in the time circle. After random average, we obtain a $Z_2$ gauge theory in the bulk. If we denote the set of all bulk links with $\sigma_{xy}=-1$ as $\Sigma$, the boundary of $\Sigma$ has to contain the future boundary $F$. If in addition to $F$ there are other boundary of $\Sigma$ in the bulk (Fig. \ref{fig:unitarity} (d)), the boundary is a $Z_2$ flux line. In the semiclassical limit, the Renyi entropy $S^{(2)}_F$ is determined by
\bea
e^{-S^{(2)}_F}=\max_{\Sigma}e^{-S^{(2)}_{\rm bulk}(\Sigma)}\label{channelcapacity}
\eea
which is determined by the bulk region $\Sigma$ that has minimal second Renyi entropy in the bulk theory.  

\addfig{5.5in}{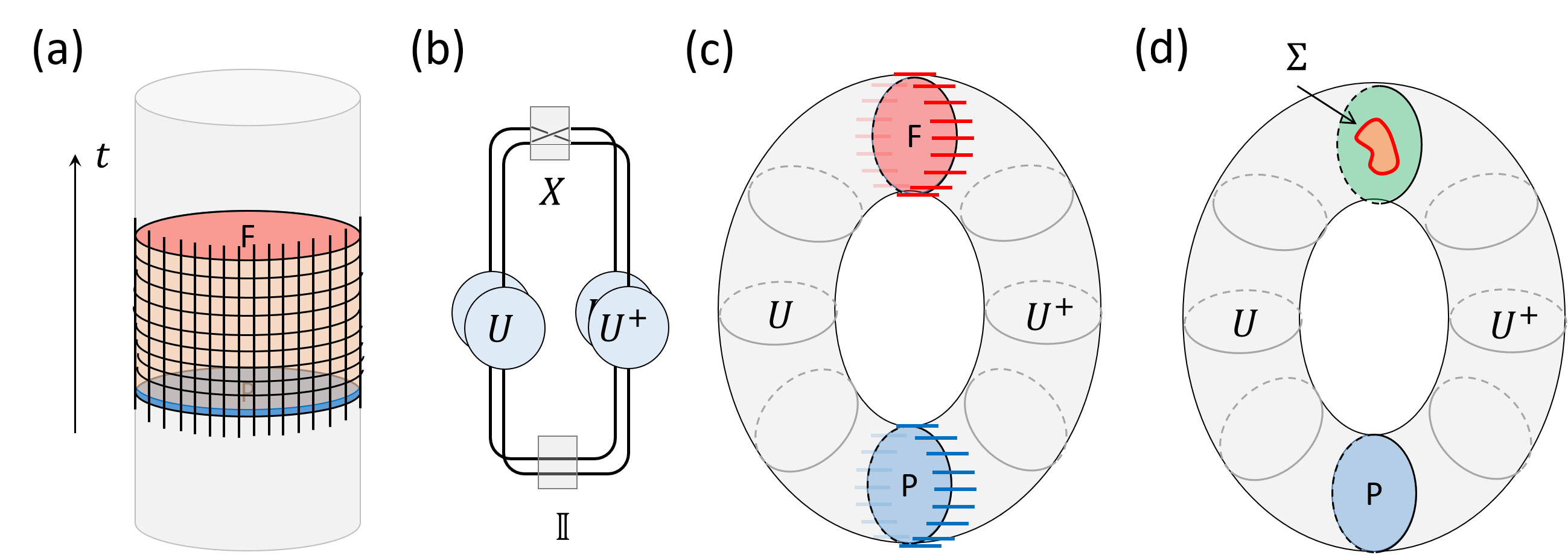}{\label{fig:unitarity} Illustration of the unitarity calculation. (a) The time evolution of the boundary in a time interval $[t_i,t_f]$ corresponds to a network with open legs on the past and future boundary, which can be viewed as a pure state in a doubled Hilbert space. The corresponding bulk geometry is a Lorentzian theory between two Cauchy surfaces $P$ and $F$ bounding the past and future boundary time slice, respectively. (b) The circuit that computes the second Renyi entropy of the future subsystem (see text). A swap operator is applied to the future subsystem of two replica, while identity operator acts on the past. (c) The bulk geometry corresponding to the circuit (b). There is a swap operator for each red link crossing the future boundary, and an identity operator for each blue link crossing the past boundary. After random average, the boundary condition of the $Z_2$ gauge theory is $\sigma_{xy}=-1$ for all red links, and $\sigma_{xy}=1$ elsewhere.  (d) Illustration of possible semiclassical configuration of the $Z_2$ gauge theory, with $\sigma_{xy}=-1$ for all links crossing the green region $\Sigma$. The bulk action determines the position of the boundary of $\Sigma$ (red closed curve). }

The channel capacity of the boundary (\ref{channelcapacity}) can have different behavior depending on the bulk theory. We will discuss two cases below.

\begin{enumerate}
\item {\bf Code subspace unitarity.} The bulk theory contains short-range UV degrees of freedom (the VBS states) and the low energy quantum field theory degrees of freedom. We can restrict the bulk time-evolution to the latter, while requiring the UV degrees of freedom to stay in its ground state. In other words, the bulk time evolution operator before random projection looks like 
\bea
U_{\rm bulk}=U_{\rm QFT}\otimes \ket{\psi^{\rm UV}_{f}}\bra{\psi^{\rm UV}_{i}}
\eea
In this case, the second Renyi entropy $S^{(2)}_{\rm bulk}(\Sigma)$ is a sum of VBS and QFT contributions:
\bea
S^{(2)}_{\rm bulk}(\Sigma)&=&\log D\abs{\partial \Sigma}+\log D_b\abs{\Sigma}\nn\\
&=&V\log D_b+\kd{\log D\abs{\partial \overline{\Sigma}}-\log D_b\abs{\overline{\Sigma}}}
\eea
Here $\overline{\Sigma}$ is the complement region of $\Sigma$, and $V$ is the net bulk volume, such that $D_b^V$ is the total dimension of the bulk low energy QFT subspace. Under the code subspace condition
\bea
\log D\abs{\partial \overline{\Sigma}}\geq\log D_b\abs{\overline{\Sigma}},~\forall \overline{\Sigma}
\eea
the minimal entropy is given by $\overline{\Sigma}=\emptyset$, which means $\Sigma$ covers the whole Cauchy surface in the bulk. It is interesting to note that the code subspace condition is the same as that in the spatial RTN\cite{hayden2016}. The second Renyi entropy in this case corresponds to 
\bea
e^{-S^{(2)}_F}=D_b^{-V}\nn
\eea
In the semiclassical limit, the calculation here can be generalized to higher Renyi entropies, and all Renyi entropies converge to the same value $V\log D_b$. Therefore the channel capacity is $I(P:F)=2V\log D_b$, which tells us that the theory has unitarity in the code subspace.
\item {\bf Unitarity in the whole boundary Hilbert space.} If we consider a large $D_b$, we can also reach the opposite situation of volume law entropy exceeding the area law entropy for all regions:
\bea
\log D\abs{\partial \overline{\Sigma}}\leq\log D_b\abs{\overline{\Sigma}},~\forall \overline{\Sigma}
\eea
In this case, the region with minimal second Renyi entropy is $\Sigma=\emptyset$, in which case the boundary $\partial \Sigma$ expands to the boundary $F$ itself. This corresponds to $S^{(2)}_F=\abs{F}\log D$, which is the maximal value of the boundary. The large $D_b$ case corresponds to a boundary-to-bulk isometry discussed in the spatial RTN case\cite{hayden2016}. Physically, such a large $D_b$ means the bulk geometry is strongly fluctuating, since, for example, different states in this big bulk Hilbert space can have completely different configurations of extremal surfaces bounding the same boundary region. 
\end{enumerate}

\section{Gauge fixing and finite D
  fluctuations}\label{sec:fluctuations}
In Sec.\ref{subsec:concept}, we point out an apparent
inconsistency in our theory, that is the boundary theory seems to be
irrelevant to the bulk tensor geometry after random projection. In contrast, as we have seen in Sec.~\ref{sec:Renyi2} to \ref{sec:operators}, the random average leads to a discrete gauge theory and gives RT formula, which clearly shows that the information of bulk geometry is encoded in the boundary theory. In this section, we will show that this apparent paradox is related to the gauge redundancy, which leads to an overall constant factor in partition function of the theory. We will show how to solve the redundancy problem by a gauge fixing, which corresponds to a refined definition of the random projections on a selected subset of links rather than all bulk links. 


\subsection{Gauge redundancy in the original  space-time tensor}\label{gauge redundancy}

In this subsection, we will first demonstrate how the inconsistency
arises, and then we will resolve the problem by gauge fixing. 

The inconsistency can be easily seen in the following calculation. For simplicity, we calculate the second Renyi entropy $\overline{e^{-S_2(A)}}$ of some boundary region $A$. In all of the previous calculations, we have assumed the following approximations. 
\begin{equation}\label{incons} \overline{\left(\frac{\tr\left[(\rho\otimes\rho)X_A\right]}{\tr[\rho\otimes\rho]}\right)} \approx \frac{\overline{\left(\tr\left[(\rho\otimes\rho)X_A\right]\right)}}{\overline{\left(\tr[\rho\otimes\rho]\right)}}
\end{equation}
However as we mentioned in Sec.\ref{subsec:concept}, if we impose random projections on all bulk links, the l.h.s of the
above equation do not depends on the bulk tensor geometry, because the
isolated bulk vertices that appear both on the denominator and the
numerator cancels before the random average. However, the r.h.s of the
above equation obviously depends on the bulk tensor network geometry,
because it calculates the free energy cost of the flux that bounds the
region A, which is dominated by the extremal surface contribution in large $D$ limit. Thus the assumption that
l.h.s and r.h.s in Eq.\ref{incons} are approximately equal to each
other is wrong. This is in sharp contrast with the situation in
spatial random tensor networks in Ref. \cite{hayden2016}, where the same approximation is
valid in the large $D$ limit. 

To analyze the origin of the problem in Eq. (\ref{incons}), we analyze the fluctuations in the random average. The l.h.s. of Eq. (\ref{incons}) can be expanded as 
\begin{eqnarray} &&\overline{\left(\frac{\tr\left[(\rho\otimes\rho)X_A\right]}
{\tr[\rho\otimes\rho]}\right)} =\overline{\left(\frac{\tr\left[(\rho\otimes\rho)X_A\right]}
{\overline{\left(\tr[\rho\otimes\rho]\right)} + \tr[\rho\otimes\rho]-\overline{\left(\tr[\rho\otimes\rho]\right)}  }\right)} \\
&=& \frac{\overline{\left(\tr\left[(\rho\otimes\rho)X_A\right]\right)}}{\overline{\left(\tr[\rho\otimes\rho]\right)}} + \frac{\overline{\left(\tr\left[(\rho\otimes\rho)X_A\right]\right)}}{\overline{\left(\tr[\rho\otimes\rho]\right)}}\cdot 
\frac{ \overline{\left(\tr[\rho\otimes\rho]\right)^2}-\left(\overline{\tr[\rho\otimes\rho]}\right)^2 }{ \left(\overline{\tr[\rho\otimes\rho]}\right)^2 } + \cdots
\end{eqnarray}
where we have only written down one single term in the leading order
of the expansion. Physically, we expect the fluctuation of $\trace{\rho\otimes \rho}$ to be small in the large bond dimension limit. However, if we literally express the random average in gauge theory partition function, on the numerator of the second term we have a term with $S^4$ gauge theory, while on the denominator we have $Z_2$ gauge theory. Note that the partition function obtained from the random average is literally a sum over gauge vector potentials, which thus contains redundant copies of the properly gauge fixed partition function. If we denote the $S_4$ gauge theory partition function with proper gauge fixing (which means only gauge inequivalent configurations are summed) as $Z_4$, and that for the $Z_2$ gauge theory as $Z_2$, in the large $D$ limit we have
\bea
\frac{\overline{\trace{\rho\otimes\rho}^2}}{\overline{\trace{\rho\otimes\rho}}^2}\simeq \kc{\frac{4!}{2!2!}}^V\frac{Z_4}{Z_2^2}
\eea
In large $D$ limit, $Z_4\simeq Z_2^2$ since they are all dominated by trivial configuration (as long as the minimal action configuration is unique modular gauge transformations). The exponentially large factor $\kc{\frac{4!}{2!2!}}^V$ is the reason of the large flucutation which makes Eq. (\ref{incons}) invalid. Our goal is to introduce a gauge fixing procedure to remove the redundancy factor. 


\subsection{Gauge fixed space-time tensor}

\begin{figure}[ht!]
  \centering
  \includegraphics[width=0.8\textwidth]{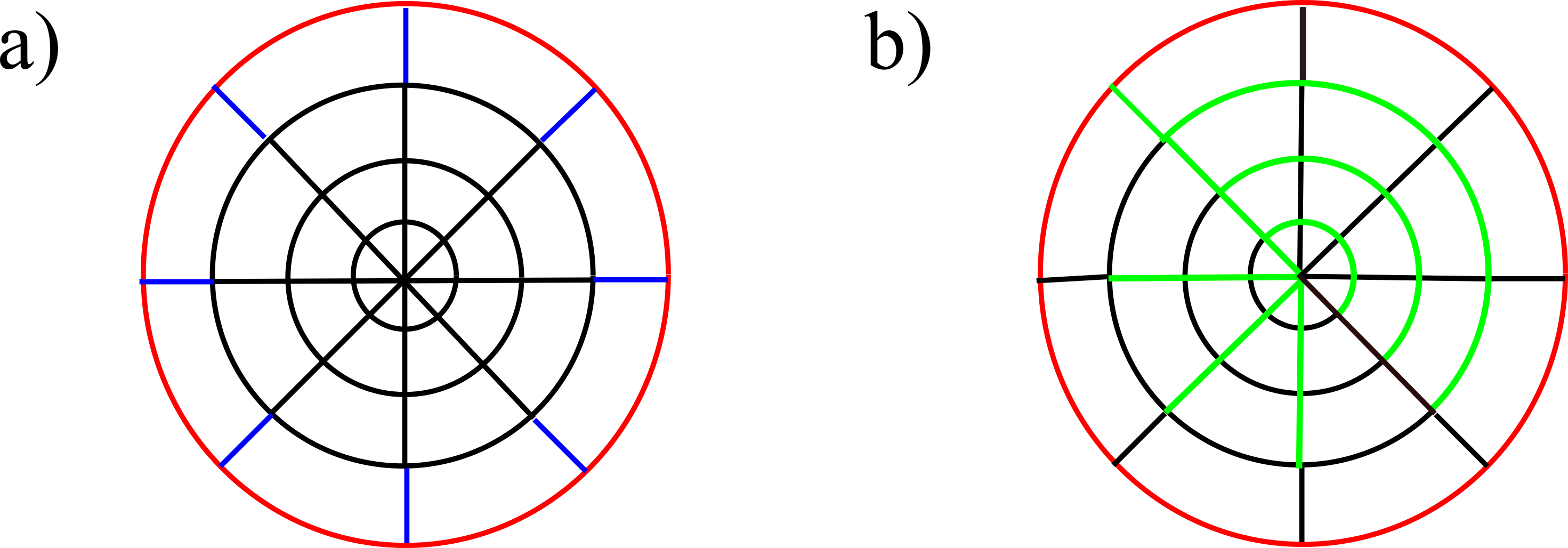}
  \caption{(a) is a tensor network in which the red tensors are the
    boundary system and the black tensors are named as \emph{isolated
      bulk tensor network}. Blue links directly connect the boundary
    system with the \emph{isolated bulk tensor network}. Fig.(b) shows
    one particular choice of the links (green links) on which the
    random projectors are removed following the gauge fixing
    procedure.}
  \label{gftensor}
\end{figure}

For discrete gauge theories, the simplest approach of gauge
  fixing is to fix the gauge vector potential on some links to
  identity. An appropriate gauge fixing is
  obtained by choosing a subset of links $F$ such that if we set the
  link variable $g_{xy}=\mathbb{I}$ for all $\avg{xy}\in F$, then 1)
  no gauge flux is fixed; 2) no gauge transformation can be carried
  while preserving the gauge fixing. These two requirements are
  equivalent to requiring $F$ to be a {\it spanning tree} of the
  network. Since the sum over permutation group element $g_{xy}$ comes
  from average over random projections on that link, the gauge fixing
  is equivalent to selecting a spanning tree $F$ and introduce random
  projections on on links that are not on $F$. More explicitly, we
  summarize the general procedure of gauge fixing below, which is also
  illustrated in Fig. \ref{gftensor}. (In this figure, the red links
  represent the boundary system and all other links are the bulk.)

\begin{enumerate}
\item Exclude the boundary tensor network (the red tensors in
  Fig.\ref{gftensor}(a)) as well as the bulk links that directly
  connect the bulk and the boundary (the blue links in
  Fig.\ref{gftensor}(a)) and we first only focus on the \emph{isolated
    bulk tensor network} (the black tensor network in
  Fig.\ref{gftensor}(a)).
\item Find a spanning tree $T$ in the \emph{isolated bulk tensor
    network}.  In graph theory, a spanning tree of an undirected graph
  $G$ is a subgraph that is a tree and includes all  the vertices
  of $G$. In addition, pick one link that directly connects the boundary with the spanning tree $T$, as is shown in Fig. \ref{gftensor} (b) by the green link connecting the boundary. Denote the union of $T$ and this extra link as $F$. 
  \item The holographic theory is defined by introducing an independent random projector on all bulk links that do not belong to $F$, {\it i. e.} all the black links in Fig. \ref{gftensor} (b). 
\end{enumerate}

From the gauge fixing procedure above one can see directly that the gauge redundancy is completely removed by this choice. In Appendix \ref{app:counting} we provide a counting argument to further verify that. 
With this modified definition of random projection, we have for example $\overline{\trace{\rho\otimes \rho}}=Z_2$ and $\overline{\trace{\rho\otimes\rho}^2}=Z_4$ with the gauge redundancy removed. Also from the definition above it is clear that the boundary theory obtained after random projection still has a nontrivial dependence on the bulk geometry and bulk theory since the bulk geometry is not completely detached. 

\subsection{Finite D fluctuation}

In this subsection, we prove that in the $D\rightarrow \infty$ limit,
the fluctuation can be bounded. The proof is largely parallel to the
discussion in Ref.\cite{hayden2016},  so we will only sketch the main steps here for completeness. 

In the derivation of Renyi entropies, we have also made the assumption by taking separate average of the numerator and denominator:
\bea
\overline{S_n(A)}&=&\frac1{1-n}\overline{\log\kc{\frac{\avg{X_{nA}\prod_{xy}P_{xy}\otimes P_{xy}}_{\rm bulk}}{\avg{\prod_{xy}P_{xy}}_{\rm bulk}^2}}}\nn\\
&\simeq &\frac1{1-n}\log\frac{\avg{X_{nA}\prod_{xy}\overline{P_{xy}\otimes P_{xy}}}_{\rm bulk}}{\avg{\prod_{xy}\overline{P_{xy}\otimes P_{xy}}}_{\rm bulk}}
\eea
Following the convention of Ref. \cite{hayden2016} we denote the numerator and denominator of the first line by $Z_{nA}$ and $Z_{n\phi}$, which are functionals of the random projectors. The large $D$ value of these partition functions are dominated by the classical configuration, which we denote by $Z_{nA}^\infty$ and $Z_{n\phi}^\infty$. The RT formula is given by
\bea
S_n^{RT}(A)=\frac1{1-n}\log\frac{Z_{nA}^\infty}{Z_{n\phi}^\infty}
\eea
At finite $D$, both the numerator and denominator are fluctuating. The finite $D$ deviation of the entropy from the RT value is given by
\bea
S_n(A)-S_n^{RT}(A)=\frac1{1-n}\kc{\log\frac{Z_{nA}}{Z_{nA}^\infty}-\log\frac{Z_{n\phi}}{{Z_{n\phi}^\infty}}}
\eea
This deviation of entropy is bounded by directly using the following result from Ref. \cite{hayden2016}:
\begin{itemize}
\item
If we find an upper bound for
\bea
\frac{\overline{Z_{nA}^2}}{e^{-\mathcal{A}^{(2n)}_A}}-1\leq f(D)
\eea
and $f(D)\ll 1$ in the large $D$ limit, then in the limit $f(D)\ll 1$, 
\bea
{\rm Prob}\kc{\abs{S_n(A)-S_n^{RT}(A)}\leq \delta}\geq 1-\frac{32}{\delta ^2}f(D)
\eea
\end{itemize}
Here the left-hand side of the equation means the probability that the entropy deviation from the RT formula value is smaller or equal to $\delta$. 

\emph{If the minimal action field configuration is unique}, then using
the fact that all other field configurations have a statistical weight
that is suppressed at least by a factor $\frac1D$, one can obtain a
universal but very loose bound $f(D)=\frac{C_T}{D}$ with $C_T$ the
total number of field configurations in the system. $C_T$ grows
exponentially with the volume of space-time, so that this bound only
leads to RT formula for exponentially large dimension $D$. However, if
we consider that the gauge theory is local and gapped, the quantity $-\log
\frac{\overline{Z_{nA}^2}}{e^{-\mathcal{A}^{(2n)}_{A}}}$ is the thermal free energy defined with respect to the zero temperature value, which for a gapped theory should satisfy 
\bea -\log
\frac{\overline{Z_{nA}^2}}{e^{-\mathcal{A}^{(2n)}_{A}}}\geq -{\rm
  const.\times} VD^{-\Delta_{2n}/2} \eea 
Here $\Delta_{2n}$ is the gap of the $S_{2n}$ gauge theory, which is an energy
scale of order one. Using this inequality one can reduce the
requirement of dimension $D$ to a power law of volume $V$.

All the discussion above are in exact parallel with the spatial random
tensor network, where the statistical model is a principle model with
isospin in the permutation group, rather than a gauge theory. The main
difference between the current situation and the spatial situation is
the nature of fluctuations away from the classical configuration. In
the spatial model, the boundary condition pins a domain wall of the
principle model, which is a co-dimension $1$ surface in
space. Therefore at large $D$ limit, the small fluctuation of this
surface occurs only along one perpendicular direction to the domain
wall. On comparison, in the space-time network the electric flux is a
co-dimension $2$ surface in the space-time, so that fluctuations
around a surface occurs along two perpendicular directions. In large
$D$ limit where $D$ is bigger than a power law of $V$, this difference
is not important since the fluctuations are small anyway. However if
$D$ is large but finite while $V\rightarrow \infty$, it is possible
that the field configuration in the bulk is still given by the minimal
co-dimension-$2$ surface plus nearby fluctuations, but it is not
dominated by any single configuration any more. This situation is an
analog of the ``roughening transition" in the case of principle
models\cite{kogut1981fluctuating, luscher1981symmetry}, but the
fluctuations have twice many degrees of freedom in the space-time case. It is an
interesting question whether the codimension difference leads to any qualitatively different behavior.

From the discussion above we see why the gauge fixing procedure is essential for bounding the fluctuation, since the minimal action configuration cannot be unique if the gauge redundancy were not removed. After gauge fixing, fluctuation is bounded as long as the minimal action gauge field configuration is uniquely determined by the boundary condition.

\section{Conclusion}\label{sec:conclusion}

In this paper, we construct a new class of models with various holographic properties using space-time tensor networks. Starting from a bulk ``parent theory" defined as a bulk space-time tensor network, introducing random projections on bulk links defines a new partition function, which can be interpreted as that of a boundary theory with a bulk dual. The bulk theory consists of UV degrees of freedom that contributes short-range area law entanglement entropy and IR degrees of freedom that describes the low energy long wavelength dynamics. The short-range entangled UV degree of freedom can be considered as ``skeleton of the bulk space-time geometry", which is responsible for the leading contribution to entanglement entropy of a boundary region. We show that the random average of replicated partition function is equivalent to that of a permutation group gauge theory. In proper large $N$ limit which corresponds to weakly coupled limit of the gauge theory, the Renyi entropies of all boundary regions satisfy the HRT formula. Our method allows to study the (random averaged) behavior of generic bulk-boundary correlation functions. Code subspace operators in the bulk (those that only applies to the low energy degrees of freedom) can be locally reconstructed on the boundary in the same way as in AdS/CFT. As an example of other quantum information properties that can be studied in random tensor networks, we discuss how our theory preserves unitarity in either the code subspace or the whole boundary Hilbert space, depending on the dimension of bulk degrees of freedom. 

Compared to spatial random tensor networks, the space-time formalism is covariant and does not rely on particular choice of time slice. A tensor network is a discretization of a path integral. Although particular discretization necessarily breaks Lorentz symmetry and other symmetries of the space-time geometry, one can consider a family of more and more refined discretization of a given space-time manifold, such that the space-time symmetries are asymptotically restored. For example, we can define a bulk parent theory that is a lattice regularization of bosons in pure AdS. In the limit of small lattice constant $a$, all correlation functions on the boundary at length scale much longer than $a$ will preserve conformal symmetry on the boundary, as a consequence of the AdS symmetry in the bulk. Naively, one would conclude that the space-time tensor networks on pure AdS geometry describes partition functions of conformal field theories. However, this is apparently inconsistent with the behavior of Renyi entropies we found. For example for CFT$_2$, we know that the Renyi entropy $S_n$ has a nontrivial $n$-dependence, while the AdS$_3$ random tensor network has all $S_n$ the same in the semi-classical limit. Similar to the spatial random tensor networks, this problem indicates that one should view the space-time random tensor networks as models with short-range gravitational interaction---Roughly speaking, it has correct degrees of freedom but it missed the gapless gravitons which mediate long range gravitational interaction in the bulk. Correspondingly, on the boundary side the problem is that these theories can have conformal invariant correlation functions (if the bulk is AdS) but it does not have a local energy momentum tensor. (For a recent discussion on such theories see Ref.~\cite{paulos2016s}.) 

There are a lot of interesting open questions. Is it possible to impose additional conditions on the space-time random tensor network such that it has a local energy-momentum tensor? In the holographic derivation of linearized Einstein equation \cite{lashkari2014gravitational,faulkner2014gravitation,swingle2014universality}, the key ingredients are the RT formula, conformal symmetry of the boundary vacuum state and the local energy momentum tensor. Since the space-time tensor newtork on AdS geometry satisfies the first two conditions, if we understand how to impose the condition of local energy momentum tensor, we may be able to understand why the bulk geometry is required to satisfy the Einstein equation. It is also possible to relate the space-time random tensor networks to another derivation of Einstein equation in Ref. \cite{jacobson2016entanglement} which does not use the boundary but starts from a hypothesis that the ``total entropy" $S_{QFT}+\frac{A}{4G}$ is extremized for small disk regions in the bulk, if we vary the geometry and preserve the volume of the disk. ($A$ is the area of the disk) In the space-time tensor networks, if we consider a flux loop along the boundary of a small bulk disk, it's free energy in the gauge theory is given by
\bea
F=(n-1)\kd{S_n^{\rm QFT}+\alpha A}
\eea
(with the area law coefficient $\alpha=\log D$ in the VBS model). Therefore maximizing $F$ corresponds to minimizing the contribution of such closed loops as fluctuations around the semi-classical saddle point. It is not clear to us how to translate this intuition to a more rigorous derivation, but this analog seems to suggest that Einstein equation may emerge from the requirement of minimizing gauge field fluctuations.

Another question is how to develop a background independent theory. In this paper we start from a parent theory in the bulk, which is already defined on a background geometry. In the case of spatial tensor networks, Ref. \cite{qi2017holographic} shows how to allow arbitrary superposition of tensor network states on different graphs by introducing link variables. It will be interesting to develop the space-time analog of that.

\noindent{\bf Acknowledgement.} 
We would like to thank Xi Dong, Patrick Hayden, Juan Maldacena, Sepehr Nezami, Mukund Rangamani, Leonard Susskind and Michael Walter for helpful discussions. This work is supported by the David and Lucile Packard Foundation. 

\appendix

\section{Analytic continuation for Relative entropy calculation}\label{app:re}

In this appendix we demonstrate why 
\begin{equation}
   \overline{\tr_B\left(\sigma_B\rho_B^{n}\right)} \simeq D^{-n|\gamma_B|}\cdot \tr_{E_B}\left(\sigma^L_{E_B}\left(\rho^L_{E_B}\right)^{n}\right)= \tr_{E_B}\left(\sigma^{bulk}_{E_B}\left(\rho^{bulk}_{E_B}\right)^{n}\right) ~~ \forall n\geq 0\label{app:sigmarho2}
\end{equation}
implies
\begin{eqnarray}\label{app:sigmarho} \tr_{B}\left(\sigma_B\log\rho_B\right)=\tr_{E_B}\left(\sigma_{E_B}^{bulk}\log\rho_{E_B}^{bulk}\right)
\end{eqnarray}

We diagonalize $\sigma_B,\rho_B,\sigma_{E_B}^{bulk},\rho_{E_B}^{bulk}$ in Eq.\ref{app:sigmarho2} to obtain
\begin{eqnarray}
 \sum_{ij} \lambda^{\sigma_B}_i \left(\lambda^{\rho_B}_j\right)^n \left(U^\dagger V\right)_{ij}\left(V^\dagger U\right)_{ji}= \sum_{ij}\lambda^{\sigma_{E_B}^{bulk}}_i \left(\lambda^{\rho_{E_B}^{bulk}}_j\right)^n \left(X^\dagger Y\right)_{ij}\left(Y^\dagger X\right)_{ji}
\end{eqnarray}
where $U,V,X,Y$ are the unitary matrices that diagonalize
$\sigma _B, \rho_B, \sigma_{E_B}^{bulk},\rho_{E_B}^{bulk}$, respectively, and $\lambda^{\sigma_B}_i,\lambda^{\rho_B}_i,\lambda^{\sigma_{E_B}^{bulk}},\lambda^{\rho_{E_B}^{bulk}}$ are the corresponding eigenvalues.  For convenience, we define $a_j = \sum_i\lambda^{\sigma_B}_i  \left(U^\dagger V\right)_{ij}\left(V^\dagger U\right)_{ji}$, $b_j =\sum_i \lambda^{\sigma_{E_B}^{bulk}}_i  \left(X^\dagger Y\right)_{ij}\left(Y^\dagger X\right)_{ji}$. Then we obtain
\begin{equation*}
  \sum_{j} \frac{a_j}{1-z\lambda^{\rho_B}_j} =\sum_{n,j} a_j z^n \left(\lambda^{\rho_B}_j\right)^n = \sum_{n,j} b_j z^n \left(\lambda^{\rho_{E_B}^{bulk}}_j\right)^n =  \sum_{j} \frac{b_j}{1-z\lambda^{\rho_{E_B}^{bulk}}_j}
\end{equation*}
The above series converge when $|z|<\min\left(\{1/\lambda^{\rho_B}_j\}, \{1/\lambda^{\rho_{E_B}^{bulk}}_j\} \right)$. Since both sides are holomorphic function and are identical in a non-empty open set, by analytical continuation, the two functions are the same. Thus they have the same poles and same residues respectively, from which we obtain 
\begin{equation*}
  \sum_j a_j \log \left(\lambda^{\rho_B}_j\right) = \sum_j b_j \log \left(\lambda^{\rho_{E_B}^{bulk}}_j\right)
\end{equation*}
which is identical to Eq.\ref{app:sigmarho}.

\section{Counting of degrees of freedom in gauge fixing}\label{app:counting}

In this appendix, we will verify that our gauge fixing procedure in Sec. \ref{sec:fluctuations} does completely remove gauge redundancies and leave only the physical degrees of freedom by counting the number of degrees of freedom.

We start by counting the number of gauge equivalent configurations
that keep the boundary gauge field unchanged in a $S^n$ gauge theory
($n$th order symmetric group). Given a configuration whose boundary
gauge fields are $\{X^\partial_i\}$ and bulk gauge fields are
$\{X^b_\alpha\}$, the number of configurations related by gauge
transformations are
$(n!)^V\cdot M\left[\{X^\partial_i, X^b_\alpha\}\right]$, where $i$
labels the boundary links and $\alpha$ labels the bulk links.
$(n!)^V$ is the number of different configurations induced by the
gauge transformations on all the bulk vertices and $V$ is the total
number of bulk vertices.
$ M\left[\{X^\partial_i, X^b_\alpha\}\right]-1$ is the number of the
gauge transformations acting homogeneously on both the bulk vertices
and boundary vertices which commute with all the boundary gauge fields
$\{X^\partial_i\}$, but transform the bulk gauge fields
$\{X^b_\alpha\}$ to a configuration that is not gauge equivalent to
itself by the gauge transformation acting on the bulk vertices.

In the large $D$ limit, the minimal action field configurations in the
$S^n$ gauge theory are those with the minimal area gauge flux. One
interesting observation is that among all the gauge equivalent
configurations, there exists one whose gauge fields in the bulk
$\{X^b_\alpha\}$ are the same as those on the boundary
$\{X^\partial_i\}$. Thus $M\left[\{X^\partial_i, X^b_j\}\right]=1$,
because all the gauge transformations acting homogeneously on the bulk
vertices and boundary vertices that commute with $\{X^\partial_i\}$
also commute with $\{X^b_\alpha\}$. Thus the gauge equivalent
configurations for the minimal action field configurations in the
$S^n$ gauge theory is $(n!)^V$. Therefore this redundancy is completely removed by our gauge fixing procedure.


\bibliographystyle{JHEP}
\bibliography{holography}

\end{document}